\documentclass{article}

\usepackage{arxiv}

\usepackage[utf8]{inputenc} 
\usepackage[T1]{fontenc}    
\usepackage[hidelinks]{hyperref}       
\usepackage{url}            
\usepackage{booktabs}       
\usepackage{amsfonts,amsmath,amssymb,bm}       
\usepackage{nicefrac}       
\usepackage{microtype}      
\usepackage{lipsum}		
\usepackage{graphicx,float,placeins}
\usepackage{doi}
\usepackage[per-mode=fraction,exponent-product=\cdot]{siunitx}
\usepackage{multirow, multicol, makecell, booktabs}
\usepackage{caption, subcaption}
\usepackage[backend=biber,style=numeric,sorting=none,sortcites,maxbibnames=99,giveninits=true]{biblatex}
\let\vec\bm

\title{A Validated Finite Element Model for Room Acoustic Treatments with Edge Absorbers}

\author{
\href{https://orcid.org/0000-0002-8652-5048}{\includegraphics[scale=0.06]{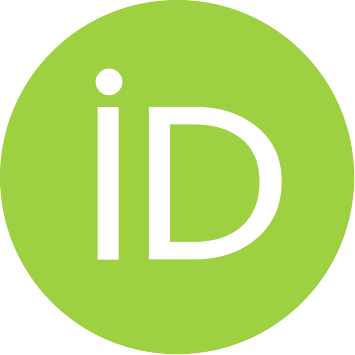}\hspace{1mm}Florian Kraxberger}\\
Institute of Fundamentals and Theory\\
in Electrical Engineering (IGTE)\\
Graz University of Technology\\
Inffeldgasse 18/I, 8010 Graz, Austria\\
\texttt{kraxberger@tugraz.at} \\
\And
\href{https://orcid.org/0000-0002-8711-9164}{\includegraphics[scale=0.06]{orcid.pdf}\hspace{1mm}Eric Kurz}\\
Signal Processing and Speech\\
Communication Laboratory (SPSC)\\
Graz University of Technology\\
Inffeldgasse 16c, 8010 Graz, Austria\\
\texttt{eric.kurz@tugraz.at} \\
\And
Werner Weselak\\
Signal Processing and Speech\\
Communication Laboratory (SPSC)\\
Graz University of Technology\\
Inffeldgasse 16c, 8010 Graz, Austria\\
\texttt{werner.weselak@tugraz.at} \\
\And
\href{https://orcid.org/0000-0002-8420-6051}{\includegraphics[scale=0.06]{orcid.pdf}\hspace{1mm}Gernot Kubin}\\
Signal Processing and Speech\\
Communication Laboratory (SPSC)\\
Graz University of Technology\\
Inffeldgasse 16c, 8010 Graz, Austria\\
\texttt{gernot.kubin@tugraz.at} \\
\And
\href{https://orcid.org/0000-0001-5511-8610}{\includegraphics[scale=0.06]{orcid.pdf}\hspace{1mm}Manfred Kaltenbacher}\\
Institute of Fundamentals and Theory\\
in Electrical Engineering (IGTE)\\
Graz University of Technology\\
Inffeldgasse 18/I, 8010 Graz, Austria\\
\texttt{manfred.kaltenbacher@tugraz.at} \\
\And
\href{https://orcid.org/0000-0002-2148-6703}{\includegraphics[scale=0.06]{orcid.pdf}\hspace{1mm}Stefan Schoder}\\
Institute of Fundamentals and Theory\\
in Electrical Engineering (IGTE)\\
Graz University of Technology\\
Inffeldgasse 18/I, 8010 Graz, Austria\\
\texttt{stefan.schoder@tugraz.at} \\
}

\renewcommand{\shorttitle}{FE Model for Room Acoustic Edge Absorbers}

\hypersetup{
pdftitle={A Validated Finite Element Model for Room Acoustic Treatments with Edge Absorbers},
pdfsubject={physics.class-ph},
pdfauthor={Florian Kraxberger, Eric Kurz, Werner Weselak, Gernot Kubin, Manfred Kaltenbacher, Stefan Schoder},
pdfkeywords={Edge Absorber, Room Acoustics, Finite Element Method, JCAL-Model, openCFS;},
}

\begin{document}
\maketitle

\begin{abstract}
Porous acoustic absorbers have excellent properties in the low-frequency range when positioned in room edges, therefore they are a common method for reducing low-frequency reverberation.
However, standard room acoustic simulation methods such as ray tracing and mirror sources are invalid for low frequencies in general which is a consequence of using geometrical methods, yielding a lack of simulation tools for these so-called edge absorbers.
In this article, a validated finite element simulation model is presented, which is able to predict the effect of an edge absorber on the acoustic field. With this model, the interaction mechanisms between room and absorber can be studied by high-resolved acoustic field visualizations in both room and absorber.
The finite element model is validated against transfer function data computed from impulse response measurements in a reverberation chamber in style of ISO 354. The absorber made of Basotect\textsuperscript{\textregistered} is modeled using the Johnson-Champoux-Allard-Lafarge model, which is fitted to impedance tube measurements using the four-microphone transfer matrix method.
It is shown that the finite element simulation model is able to predict the influence of different edge absorber configurations on the measured transfer functions to a high degree of accuracy. The evaluated third-octave band error exhibits deviations of \SI{3.25}{\decibel} to \SI{4.11}{\decibel} computed from third-octave band averaged spectra.
\end{abstract}


\section{Introduction}
How humans perceive the acoustic quality of a room is one of the key usability criteria in carefully designed buildings. For example, the indoor sound environment has an influence on human task performance \cite{Reinten2017indoor}. Controlling the acoustical properties of a room is of utmost importance for the perceived comfort in the room because negatively perceived acoustic environments lead to increased distraction, concentration difficulties and reduced privacy \cite{KaarlelaTuomaala2009Effects}. The acoustical quality, together with the visual quality, was found to be highly relevant for school performance due to the influence on speech comprehension for elementary-school teachers \cite{Levandoski2022Quality} as well as for secondary-schools students \cite{Astolfi2008Subjective}. Furthermore, a necessity of well-planned acoustic measures for rooms is given by the fact that noise level is negatively correlated with job satisfaction \cite{Park2020Associations,Loh2019Objective}.

A conventional approach to predicting the acoustic properties of a room is using geometrical acoustics, i.e., ray tracing simulations and mirror sources \cite{Savioja2015Overview,Vorlaender2013Computer}.
These methods can deliver accurate results for higher frequencies, although they neglect wave phenomena such as diffraction, by assuming sound to propagate as rays. However, for low frequencies $f$ and/or small geometric dimensions $d$ (i.e., for low Helmholtz numbers $He = 2\pi f d / c \ll 1$), these methods deliver incorrect results as the assumptions of ray tracing and mirror source methods are not fulfilled \cite{Savioja2015Overview}.
At the same time, low frequencies often present a problem in room acoustic scenarios, due to long multi-exponential decay of low-frequency (non-diffuse) room modes, which is present when evaluating, e.g., third-octave bands \cite{Prato2016Reverberation, Balint2018Multi}. The reason for the multi-exponential decay can be found in interference and low modal density (i.e., few modes per frequency interval) \cite{Oelmann1986Messung}. To damp these unwanted modal reverberation tails, one acoustic treatment method employs porous absorbent material placed in or close to the edges of the room, which are hence called "bass traps" or edge absorbers (EA) \cite{Fuchs2013Covered,Kurz2020Edge}. This has already been observed by Maa in 1940, who concluded regarding absorber placement, that \textit{"the most effective positions are along the edges and especially at the corners"} \cite[p.~51]{Maa1940Non‐Uniform}, because of the acoustic \textit{pressure} maxima.
However, in the case of porous absorbers, kinetic energy, related to the acoustic \textit{particle velocity}, is transformed to heat and thus extracted from the sound field. This effect originates from the friction losses due to the particle movement of the air in the viscous boundary layer in the absorber's pores \cite[Ch.~6.4]{Moeser2009Engineering}.
Since the acoustic particle velocity in the corner of a room is forced to zero, the positioning of porous absorbent material in the edge is preferable over the corner in order to damp room modes that are pronounced along the edge absorber length \cite{Kurz2022Interference}. This placement strategy was supported by theoretical investigations of Waterhouse, who described the sound field interference patterns in front of an infinite acoustically rigid edge analytically assuming infinitely many diffusely (i.e., isotropically and incoherently) incident planar sound waves \cite{Waterhouse1955Interference}. However, as shown in \cite{Kurz2022Interference}, these interference patterns can be confirmed by measurement in a reverberation chamber (RC) only for frequencies whose wavelengths are smaller than approximately 1/3 of the longest room dimension (in \cite{Kurz2022Interference} this results in $f \geq \SI{125}{\hertz}$).

Considering that conventional room acoustic simulation methods (such as ray tracing and mirror sources) are not able to simulate the low-frequency range, where the wavelengths of the considered sound waves are in the same scale as the room dimensions, they cannot be used to simulate the influence of edge absorbers for low frequencies. Thus, numerical wave-based methods must be employed to properly model wave phenomena and the influence of edge absorbers. The Finite Element Method (FEM) presents an appropriate tool and has been used to model low-frequency room acoustics, as the following examples illustrate:
In \cite{Sevastiadis2010Analysis}, a FEM simulation of a small studio is presented using a commercial FEM code where acoustic absorbers were modeled using surface impedances.
A two-dimensional FEM for modeling the acoustic effect of micro-perforated panels has been presented in \cite{Okuzono2015finite}, where a dedicated finite element type was introduced for modeling micro-perforated panels. This approach was extended to the acoustic field in rooms in three dimensions in \cite{Okuzono2018frequency}.
Regarding time-domain FEM, a comparison between explicit and implicit methods for room acoustics has been presented in \cite{Okuzono2016explicit} using frequency-independent finite impedance boundary conditions, and the explicit solver was recently extended with locally reacting frequency-dependent impedance boundary conditions  and has been applied to a large-scale auditorium \cite{Yoshida2022Parallel}. Recently, this model was used to investigate different planar absorber configurations in a room \cite{Okuzono2022time}.
However, to the authors' best knowledge, there is currently no validated FEM able to predict the influence of volumetric porous EAs with an equivalent fluid model on the acoustic field on a room, which is the aim of this paper.

In the present work, the homogeneous Helmholtz equation solved by the FEM implemented in the open-source FE-framework \textit{openCFS} \cite{Schoder2022openCFS} is used to model the influence of different EA configurations on the acoustic pressure field in a reverberant room. For parametrizing the Helmholtz equation, the so-called equivalent fluid model approach is used \cite{Kaltenbacher2018Nonconforming} with the fitting procedure published in \cite{Floss2021Design,Floss2022Mitigation}. The FE model is verified by means of a mesh convergence study and validated against impulse response (IR) measurements in the RC at TU Graz. The measurements used for validation are documented in \cite{Hofer2022Analyse}. The new model addresses the lack of simulation models for porous room acoustic EAs, which has been discussed in \cite{Kurz2020Edge}.

The paper is organized as follows. Section~\ref{sec:FEmodel} presents the numerical model used to simulate the room acoustic properties with and without different EA configurations. In section~\ref{sec:validation}, the validation procedure containing a grid study and comparisons to measurements is described. The field results of the simulation model are presented in section~\ref{sec:results}. Finally, section~\ref{sec:conclusion} discusses the main findings.

\section{Simulation Model}\label{sec:FEmodel}

\subsection{Room Specifications and Absorber Configurations}
The simulation model represents the cuboid RC of the Laboratory for Building Physics at Graz University of Technology \cite{LFB2023Hallraum}, which was emptied from any resonators and diffusers.
The RC walls are made of steel reinforced concrete and the edge lengths are $l_x=\SI{8.34}{\meter}$, $l_y=\SI{5.99}{\meter}$, and $l_z=\SI{4.90}{\meter}$.
The lower frequency limit for the gradually beginning build-up of a homogeneous sound field in the RC can be derived via the mode density $\Delta N/\Delta f$. 
With the extended equation for the number of mode frequencies up to a certain frequency $f$ \cite[eq.~(3.20a)]{Kuttruff2009RoomAcoustics} and a relative bandwidth $\Delta f/f$, $\Delta N$ is
\begin{equation}
    \Delta N = \frac{\Delta f}{f} \left( 4 \pi V \left( \frac{f}{c} \right)^3 + \frac{\pi}{2} S \left( \frac{f}{c} \right)^2 + \frac{1}{8} L \frac{f}{c} \right) \, ,
    \label{eq:DeltaN}
\end{equation}
where $c$ is the speed of sound, $V$ is the volume, $S$ is the total area of all surfaces, and $L$ is the total edge length of the RC.
If, in style of \cite{Oelmann1986Messung}, a mode number of $\Delta N = 30$ is assumed in a one-third octave band with relative bandwidth $\Delta f/f = \sqrt[3]{2} -1$ around the center frequency $f$ and eq.~(\ref{eq:DeltaN}) is rearranged, a lower frequency limit of $f \approx \SI{100}{\hertz}$ is obtained for the RC. 
The sound field in RC for the frequency range below $f$ is fully dominated by room modes. 

\begin{figure}[htbp]
    \centering
    \begin{subfigure}[b]{0.5\textwidth}
        \centering
        \includegraphics[scale=0.7]{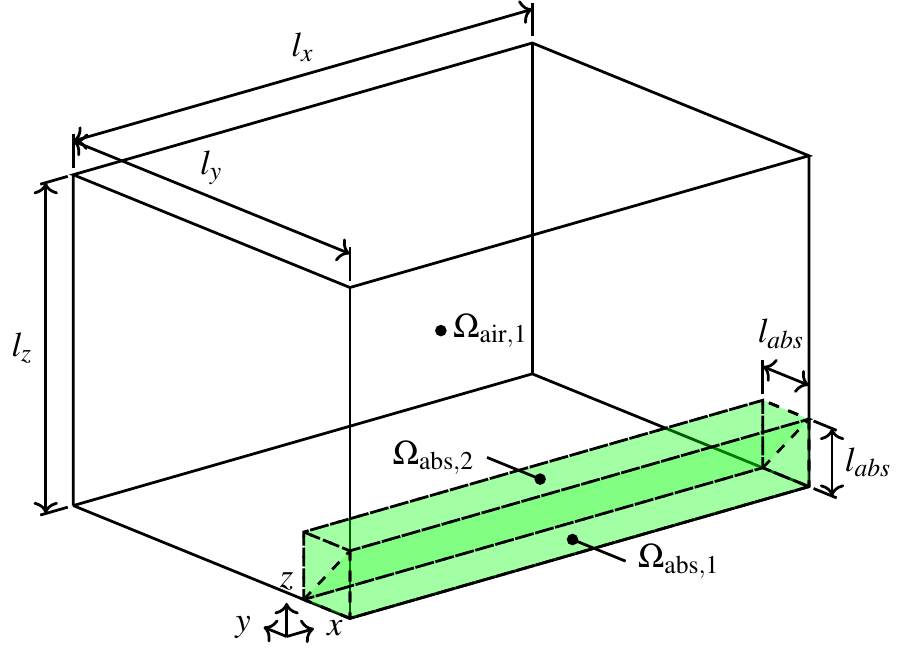}
        \caption{Sketch of geometry (not to scale). The volumes making up the absorbent volume are colored greenish.}
        \label{fig:geometry-sketch}
    \end{subfigure}
    \hfill
    \begin{subfigure}[b]{0.4\textwidth}
        \centering
        \includegraphics[scale=0.7]{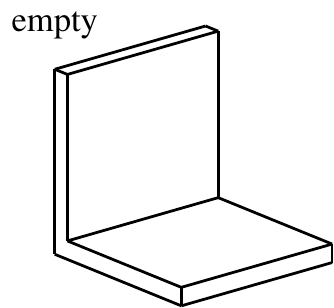}
	    \includegraphics[scale=0.7]{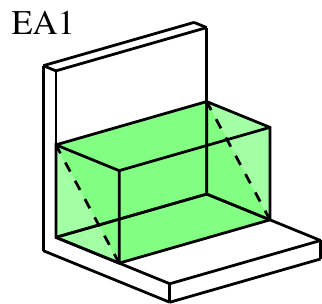}
	
	    \includegraphics[scale=0.7]{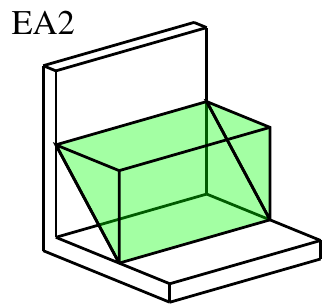}
	    \includegraphics[scale=0.7]{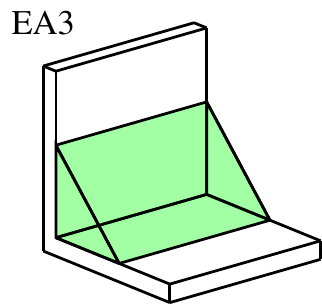}
        \caption{Simulation configurations empty, EA1, EA2, and EA3. The absorbent material is colored greenish.}
        \label{fig:absorber-configs}
    \end{subfigure}
    \caption{Geometry sketch and different absorber configurations}
    \label{fig:geometry}
\end{figure}

The porous absorber material is a melamine resin foam made of Basotect\textsuperscript{\textregistered} with a material density of $\rho_A = \SI{9}{\kilogram\per\meter^3}$ and a length-specific flow resistance of $\sigma = \SI{13}{\kilo\newton\second\per\meter^4}$ (specifications from the manufacturer).
The geometric setup of the different EA configurations is sketched in fig.~\ref{fig:geometry}. The EA with a side length of $l_\mathrm{abs}=\SI{0.4}{\meter}$ is placed along the $x$-axis in the room edge as indicated in fig.~\ref{fig:geometry-sketch}, which is the largest room dimension.

The volumes $\Omega_{\mathrm{air},1}$, $\Omega_{\text{abs},1}$ and $\Omega_{\text{abs},2}$ are now filled virtually with either air or absorber material according to the simulation configuration, i.e., they are combined into the two material volumes $\Omega_{\mathrm{air}}$ and $\Omega_{\text{abs}}$ according to table~\ref{tab:material-regions}. The resulting absorbent volume is colored greenish in fig.~\ref{fig:absorber-configs}.

\begin{table}[htbp]
    \centering
    \caption{Definitions of simulation configurations.}
    \begin{tabular}{@{}lll@{}}
        \toprule
        configuration & 
        air volume $\Omega_\mathrm{air}$ & 
        absorber volume $\Omega_\mathrm{abs}$ \\
        \midrule
        empty  & $\Omega_\mathrm{air}=\Omega_{\mathrm{air},1} \cup \Omega_{\text{abs},1} \cup \Omega_{\text{abs},2}$ & $\Omega_\mathrm{abs} = \{ \}$  \\
        EA1  & $\Omega_\mathrm{air}=\Omega_{\mathrm{air},1}$  & $\Omega_\mathrm{abs} = \Omega_{\text{abs},1} \cup \Omega_{\text{abs},2}$  \\
        EA2  & $\Omega_\mathrm{air}=\Omega_{\mathrm{air},1} \cup \Omega_{\text{abs},1}$  & $\Omega_\mathrm{abs} = \Omega_{\text{abs},2}$ \\
        EA3  & $\Omega_\mathrm{air}=\Omega_{\mathrm{air},1} \cup \Omega_{\text{abs},2}$  & $\Omega_\mathrm{abs} = \Omega_{\text{abs},1}$\\
        \bottomrule
    \end{tabular}
    \label{tab:material-regions}
\end{table}

For an empty cuboid room enclosed with sound hard walls, modal frequencies $f_{\mathrm{mode,analyt}}^{(n_x,n_y,n_z)}$ can be computed analytically with \cite[p.~220]{Moeser2009Engineering}
\begin{equation}
    f_{\mathrm{mode,analyt}}^{(n_x,n_y,n_z)} = \frac{c}{2}\sqrt{\left(\frac{n_x}{l_x} \right)^2 + \left(\frac{n_y}{l_y} \right)^2 + \left(\frac{n_z}{l_z} \right)^2} \, ,
    \label{eq:analyt-modes}
\end{equation}
where $n_x$, $n_y$, and $n_z$ are the mode orders, and $l_x$, $l_y$ and $l_z$ are the room's dimensions in $x$, $y$ and $z$ directions, respectively. The speed of sound $c$ can be computed from the bulk modulus $K$ and the density $\rho$ of the material, i.e. $c = \sqrt{K / \rho}$.

\subsection{Equivalent Fluid Model}
To model the acoustic wave inside the absorber volume $\Omega_\text{abs}$, eq.~\eqref{eq:helmholtz-efm} is solved in the computational domain $\Omega=\Omega_\text{abs} \cup \Omega_\text{air}$,
\begin{equation}
\begin{split}
    \frac{\omega^2}{K(\omega,\vec{x})} p(\omega,\vec{x}) + \nabla \cdot \left( \frac{1}{\rho(\omega,\vec{x})} \nabla  p (\omega,\vec{x}) \right) &= 0 \qquad \text{for} \qquad \vec{x} \in \Omega \, , \\
    \text{pressure excitation enforced by a dirichlet BC} &\quad p = \SI{1}{\pascal}  \quad \text{for} \quad \vec{x}=\vec{x}_\text{src} \, , \\
    \text{and the Neumann boundary condition} &\quad \nabla p \cdot \vec{n}=0 \quad \text{for} \quad \vec{x} \in \partial\Omega=\Gamma \, ,
\end{split}
\label{eq:helmholtz-efm}
\end{equation}
where $\omega=2\pi f$ is the angular frequency, $\vec{x}$ is a point in the computational domain $\Omega$, $\Gamma=\partial\Omega$ is the boundary of $\Omega$, $\vec{n}$ is the outward pointing normal vector of $\partial\Omega$, $\vec{x}_\text{src}$ is the source position, and $p(\omega,\vec{x})$ is the acoustic pressure, i.e. the solution quantity. The bulk modulus $K(\omega,\vec{x})$ is defined as
\begin{equation}
\label{eq:K}
    K(\omega,\vec{x}) = \begin{cases}
    K_\text{air} & \text{for } \vec{x} \in \Omega_\text{air} \\
    K_\text{abs}(\omega) & \text{for } \vec{x} \in \Omega_\text{abs}
    \end{cases},
\end{equation}
and the density $\rho(\omega,\vec{x})$ is defined as
\begin{equation}
\label{eq:rho}
    \rho(\omega,\vec{x}) = \begin{cases}
    \rho_\text{air} & \text{for } \vec{x} \in \Omega_\text{air} \\
    \rho_\text{abs}(\omega) & \text{for } \vec{x} \in \Omega_\text{abs}
    \end{cases}.
\end{equation}
Note that $K_\text{air}=\SI{141855}{\newton\per\square\meter}$ and $\rho_\text{air}=\SI{1.2305}{\kilo\gram\per\cubic\meter}$ are the bulk modulus and density of air at $\vartheta = \SI{13.6}{\celsius}$, respectively, which is the mean ambient temperature of the validation measurements (see section~\ref{sec:validation}) having a standard deviation of \SI{0.2}{\celsius}. For $K_\text{abs}(\omega)$ and $\rho_\text{abs}(\omega)$, a material model is necessary, as introduced in the following.

\subsection{JCAL Model for Porous Materials}
For determining the frequency-dependent and complex-valued \textit{equivalent bulk modulus} $K_\text{abs}(\omega)$ and \textit{equivalent fluid density} $\rho_\text{abs}(\omega)$ in the absorber volume $\Omega_\text{abs}$, the Johnson-Champoux-Allard-Lafarge (JCAL) model is used \cite{Champoux1991Dynamic,Johnson1987Theory,Lafarge1997Dynamic}. In this model, thermal damping effects are modeled by the complex-valued bulk modulus $K_\text{abs}(\omega)$, and visco-inertial damping effects by the complex-valued equivalent density $\rho_\text{abs}(\omega)$. $K_\text{abs}(\omega)$ and $\rho_\text{abs}(\omega)$ are calculated (with $\mathrm{j}$ being the imaginary unit) from
\begin{equation}
\begin{split}
    \rho_\text{abs}(\omega) &= \frac{\alpha_\infty \rho_\text{air}}{\phi} \left[ 1 + \frac{\sigma \phi}{\mathrm{j}\omega \rho_\text{air} \alpha_\infty} \sqrt{1 + \mathrm{j}\frac{4 \alpha_\infty^2 \eta_0 \rho_\text{air} \omega}{\sigma^2 \Lambda^2 \phi^2}} \right] \, ,\\
    K_\text{abs}(\omega) &= \frac{\gamma p_0 / \phi} {\gamma - (\gamma -1)\left[ 1 - \mathrm{j} \frac{\phi \kappa}{k'_0 C_\mathrm{p} \rho_\text{air} \omega  }  \sqrt{1 + \mathrm{j}\frac{4 k'^2_0 C_\mathrm{p}\rho_\text{air} \omega }{\kappa \Lambda'^2 \phi^2} }    \right]^{-1} } \, ,
\end{split}
\end{equation}
with the open porosity $\phi$, the static airflow resistance $\sigma$, the high-frequency limit of the tortuosity $\alpha_\infty$, the viscous characteristic length $\Lambda$, the thermal characteristic length $\Lambda'$, and the static thermal permeability $k'_0$. These are the six parameters of the JCAL model formulated in the parameter vector $   \vec{\theta}_\mathrm{JCAL} = \begin{bmatrix} \phi & k_0' & \Lambda & \Lambda' & \sigma & \alpha_{\infty} \end{bmatrix} ^\mathrm{T}$.
Note, that the JCAL parameters $\vec{\theta}_\mathrm{JCAL}$ are \textit{frequency-independent}.
Furthermore, the constitutive parameters of air are the dynamic viscosity $\eta_0$, thermal conductivity $\kappa$, isentropic exponent $\gamma$, the ambient air pressure $p_0$, and the specific heat of air at constant ambient pressure $C_\mathrm{p}$, for which the following values have been used, corresponding to the ambient conditions at the measurement temperature:
\begin{equation}
\begin{gathered}
    \eta_0=\SI{18.232e-6}{\kg\per\meter\per\second} \, , \qquad \kappa=\SI{25.684e-3}{\watt\per\meter\per\kelvin} \, , \qquad \gamma=1.4 \, , \\
    \qquad C_\mathrm{p}=\SI{1006.825}{\joule\per\kg\per\kelvin} \, , \qquad p_0=\SI{100325}{\pascal}
\end{gathered}
\end{equation}

From the equivalent density and bulk modulus, the characteristic impedance $Z_\mathrm{JCAL}(\omega)$, the complex wave number $k_\mathrm{JCAL}$ in the porous material, and successively the reflection coefficient $r_\mathrm{JCAL}(\omega)$ can be computed with
\begin{equation}
\begin{split}
    Z_\mathrm{JCAL}(\omega) &= \sqrt{\rho_\text{abs}(\omega) K_\text{abs}(\omega)}
    \, , \qquad k_\mathrm{JCAL}(\omega) = \omega \sqrt{\frac{\rho_\mathrm{abs}(\omega)}{K_\mathrm{abs}(\omega)}} , \\
    \vec{T}_\mathrm{JCAL} &= \begin{bmatrix}
        T_{11} & T_{12}  \\
        T_{21} & T_{22}
    \end{bmatrix} = \begin{bmatrix}
        \cos\!\left(k_\mathrm{JCAL} h \right) & \mathrm{j}Z_\mathrm{JCAL}\sin\!\left( k_\mathrm{JCAL} h \right) \\
        \mathrm{j} \frac{1}{Z_\mathrm{JCAL}} \sin\!\left( k_\mathrm{JCAL} h \right) & \cos\!\left(k_\mathrm{JCAL} h \right)
    \end{bmatrix} \, ,  \\
    r_\mathrm{JCAL}(\omega) &= \frac{T_{11}-T_{21}Z_\text{air}}{T_{11}+T_{21}Z_\text{air}} \, , 
\end{split}
\end{equation}
where $Z_\text{air}=\sqrt{\rho_\text{air} K_\text{air}}$ is the specific acoustic impedance of air, and $h=\SI{15.5}{\centi\meter}$ is the material thickness of the probe in the measurement tube. For the reflection coefficient $r_\mathrm{JCAL}(\omega)$, a sound-hard backing surface is assumed.

\begin{figure}[htbp]
    \centering
    \includegraphics[width=0.5\textwidth]{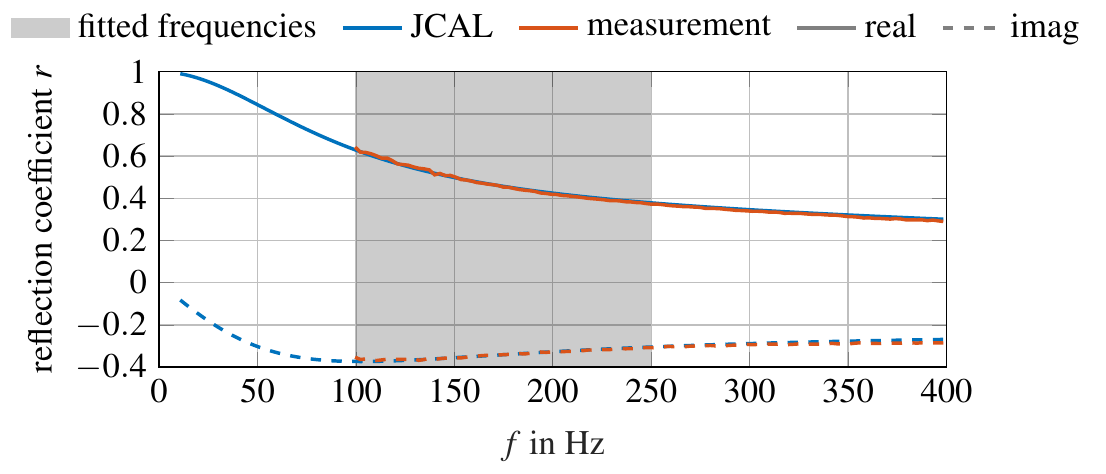}
    \caption{Measured $r_\mathrm{meas}(f)$ and fitted $r_\mathrm{JCAL}(f)$ reflection coefficient.}
    \label{fig:reflection-coeff}
\end{figure}

To obtain the parameters $\vec{\theta}_\mathrm{JCAL}$ of the JCAL model, a genetic optimization algorithm \cite{MathWorks2022GA} is used, such that the reflection coefficient $r_\mathrm{JCAL}(\omega)$ corresponds to the reflection coefficient $r_{\mathrm{meas}}(\omega)$ from impedance tube measurements with the four-microphone method (\!\textit{Transfer Matrix Method} \cite{Song2000transfer}
), as formulated in the cost function $J\!(\vec{\theta}_\mathrm{JCAL})$ in eq.~\eqref{eq:JCAL-objective}. The same procedure has been used by Floss et~al. (cf., e.g., \cite{Floss2021Design} and \cite[Ch.~3.1]{Floss2022Mitigation}), albeit they used the two-microphone method in contrast to the four-microphone method for measuring $r_{\mathrm{meas}}(f)$. The initial values for the genetic algorithm are obtained from the study of \cite[Tab.~1]{Li2020Diffuse} (Basotect\textsuperscript{\textregistered} G+) and are listed in tab.~\ref{tab:JCAL-parameters}. After initialization, the genetic algorithm \cite{MathWorks2022GA} minimizes the cost-function $J\!(\vec{\theta}_\mathrm{JCAL})$ 
\begin{equation}
\label{eq:JCAL-objective}
    J\!\left(\vec{\theta}_\mathrm{JCAL}\right) = \sum_{f=\SI{100}{\hertz}}^{\SI{250}{\hertz}} \left\lVert r_{\mathrm{meas}}(f) - r_{\mathrm{JCAL}}(\vec{\theta}_\mathrm{JCAL},f) \right\rVert_2 \, ,
\end{equation}
by adjusting the parameters $\vec{\theta}_\mathrm{JCAL}$ of the JCAL model within the bounds determined by $\vec{\theta}_\mathrm{JCAL,init}\pm 0.5\vec{\theta}_\mathrm{JCAL,init}$. Initial parameters $\vec{\theta}_\mathrm{JCAL,init}$ and final (optimal) parameters $\vec{\theta}_\mathrm{JCAL,opt}$ are listed in tab.~\ref{tab:JCAL-parameters}.
The obtained value for $\sigma$ by the fitting algorithm is in very good agreement with the manufacturer supplied value. 
The measured $r_\mathrm{meas}(f)$ and fitted (i.e. optimized) $r_\mathrm{JCAL}(f)$ reflection coefficients are depicted in fig.~\ref{fig:reflection-coeff}.
The measurements are particularly reliable for frequencies above \SI{100}{\hertz}, therefore the frequency range of the JCAL parameter starts from \SI{100}{\hertz}.
\begin{table}[htbp]
    \centering
    \caption{Initial parameters $\vec{\theta}_\mathrm{JCAL,init}$ and optimized parameters $\vec{\theta}_\mathrm{JCAL,opt}$ as a result of the fitting algorithm described in \cite{Floss2021Design,Floss2022Mitigation}.}
    \begin{tabular}{@{}llrrrlll@{}}
    \toprule
          & \multicolumn{1}{c}{$\phi$} &
          \multicolumn{1}{c}{$k_\mathrm{0}'$} &  
          \multicolumn{1}{c}{$\Lambda$} & 
          \multicolumn{1}{c}{$\Lambda'$} & 
          \multicolumn{1}{c}{$\sigma$} & 
          \multicolumn{1}{c}{$\alpha_{\infty}$} & reference\\
        \midrule
       $\vec{\theta}_\mathrm{JCAL,init}$  &  $0.994$ & $\SI{27e-10}{\meter^2}$ & $\SI{92}{\micro\meter}$& $\SI{197}{\micro\meter}$ & $\SI{10934}{\newton \second \per \meter^4}$ & $1.04$ & \cite[Table~1]{Li2020Diffuse}\\
       $\vec{\theta}_\mathrm{JCAL,opt}$  & $0.96548$& $\SI{39.52e-10}{\meter^2}$ & $\SI{125.8}{\micro\meter}$ & $\SI{284.4}{\micro\meter}$ & $\SI{12844}{\newton \second \per \meter^4}$ & $0.8304$ & \\
       \bottomrule
    \end{tabular}
    \label{tab:JCAL-parameters}
\end{table}

\subsection{Finite Element Model}
The weak form of the homogeneous Helmholtz equation is obtained by multiplying eq.~\eqref{eq:helmholtz-efm} with the test function $w\in H_0^1$ in the Sobolev space $H_0^1$, and integrating over the computational domain $\Omega$. After integration by parts, we arrive at the weak form
\begin{equation}
    \label{eq:EFM-weak-form}
    \int_{\Omega} w \frac{\omega^2}{K(\omega,\vec{x})} p \mathrm{d} \Omega  -  \int_{\Omega}   \frac{1}{\rho(\omega,\vec{x})} \nabla p \cdot \nabla  w \mathrm{d} \Omega  = -\int_{\Gamma} w \frac{1}{\rho(\omega,\vec{x})}   \underbrace{\nabla p \cdot \vec{n}}_\mathrm{=0} \mathrm{d} \Gamma =0  \, ,
\end{equation}
where $p\in H_0^1$ is the pressure, $\omega \in \mathbb{R}$ is the angular frequency, $\vec{x}\in\Omega$ is a point in the computational domain $\Omega$, $\Gamma = \partial \Omega$ is the boundary of $\Omega$ and $K(\omega,\vec{x})\in\mathbb{C}$ and $\rho(\omega,\vec{x})\in\mathbb{C}$ are the (equivalent) bulk modulus and density, as defined in eqs.~\eqref{eq:K} and \eqref{eq:rho}.
Note that the right-hand side is zero due to the homogeneous Neumann boundary condition of sound-hard walls. Equation~\ref{eq:EFM-weak-form} is spatially discretized using second-order Lagrangian finite elements. A non-uniform grid has been used, as depicted in fig.~\ref{fig:convergence-meshes}, for which the FE formulation with Nitsche-type mortaring has been introduced in \cite{Kaltenbacher2018Nonconforming} and implemented in the open-source FEM solver \textit{openCFS} \cite{Schoder2022openCFS}. The frequency-domain FEM system is solved with respect to the boundary conditions as introduced in eq.~\eqref{eq:helmholtz-efm}. To this end, the frequency range from \SI{0}{\hertz} to \SI{200}{\hertz} is quantized with frequency steps of \SI{0.5}{\hertz}. 

\section{Validation Procedure}\label{sec:validation}

The validation procedure of the FE simulation setup consists of a mesh-convergence study in order to verify the convergence of the FE model \cite{Schoder2020Hybrid}, and a point-wise comparison of the FE simulation results to transfer functions (TFs) obtained from IR measurements in the RC.

\subsection{Verification --- Convergence Study}

\begin{figure}[htbp]
    \centering
    \begin{subfigure}[b]{0.25\textwidth}
        \centering
        \includegraphics[width=\textwidth,keepaspectratio,trim=0cm 0cm 0cm 0cm,clip]{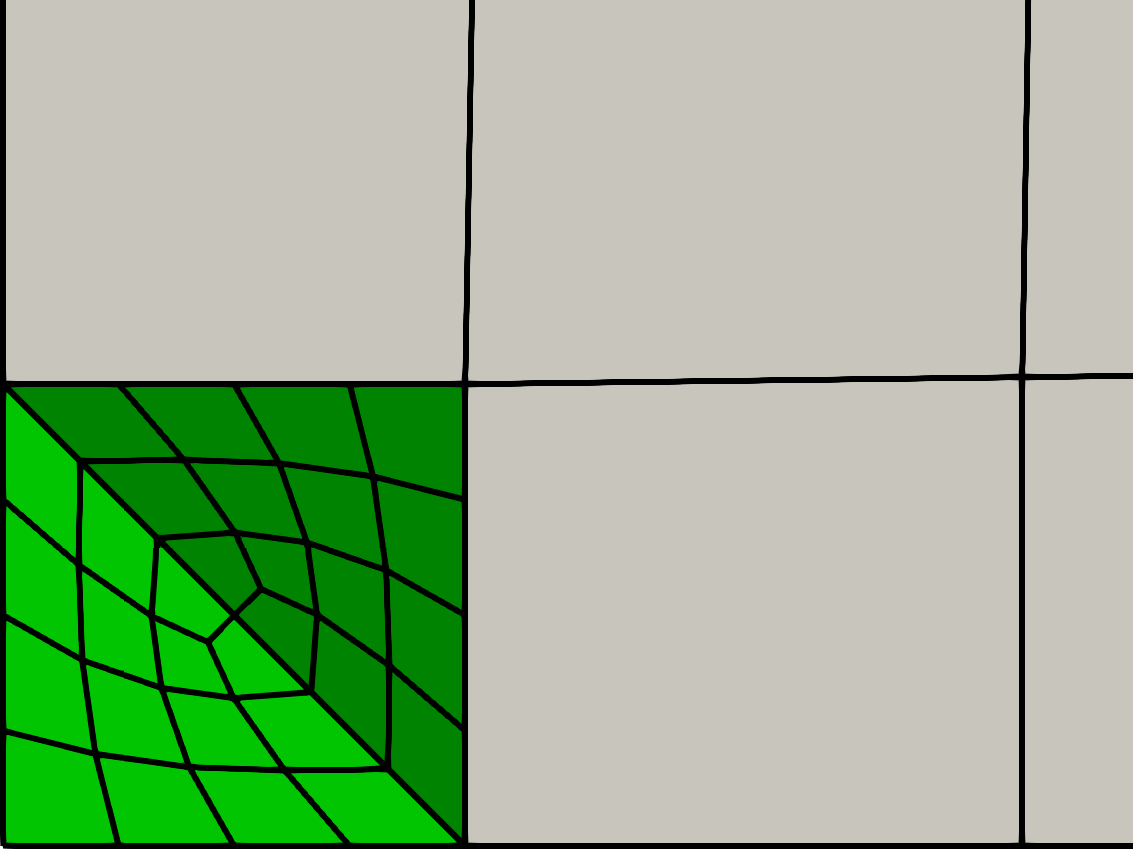}
        \caption{$\lambda/3$ mesh.}
        \label{fig:mesh-lambda3}
    \end{subfigure}
    \hspace{1cm}
    \begin{subfigure}[b]{0.25\textwidth}
        \centering
        \includegraphics[width=\textwidth,keepaspectratio,trim=0cm 0cm 0cm 0cm,clip]{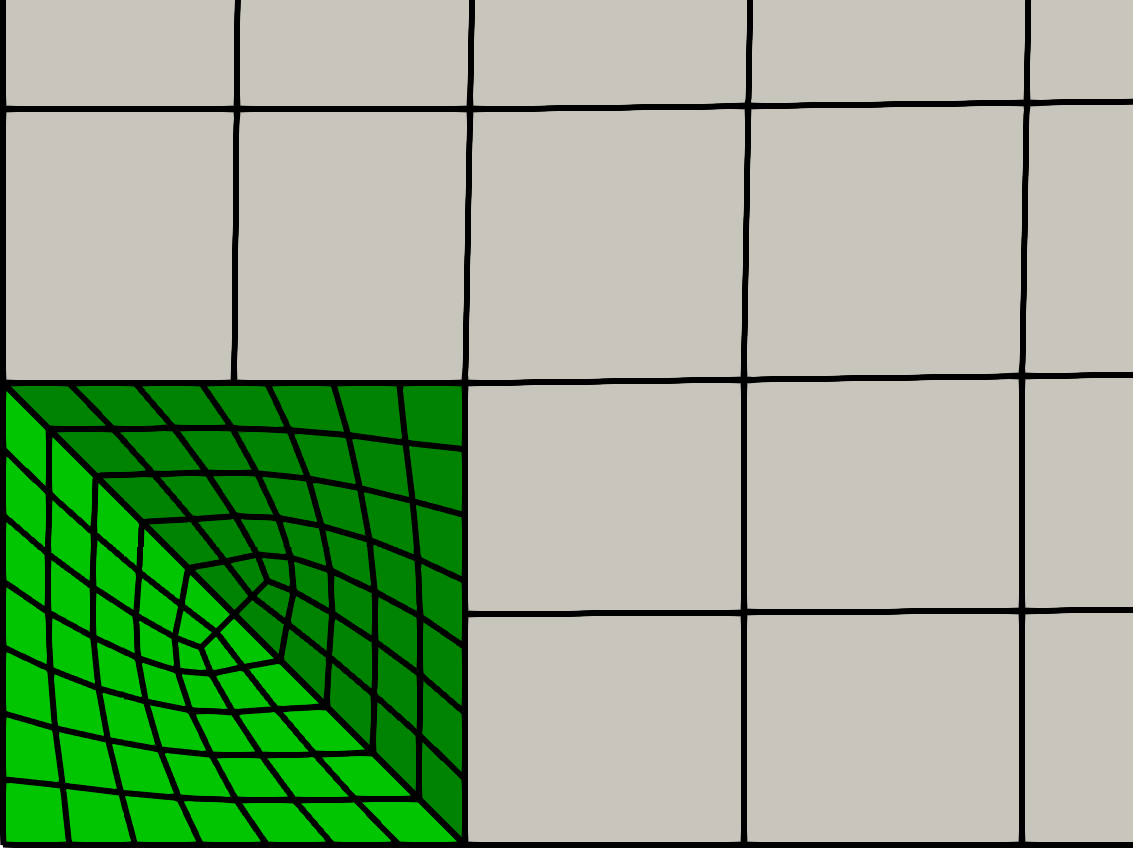}
        \caption{$\lambda/6$ mesh.}
        \label{fig:mesh-lambda6}
    \end{subfigure}\\
    \begin{subfigure}[b]{0.25\textwidth}
        \centering
        \includegraphics[width=\textwidth,keepaspectratio,trim=0cm 0cm 0cm 0cm,clip]{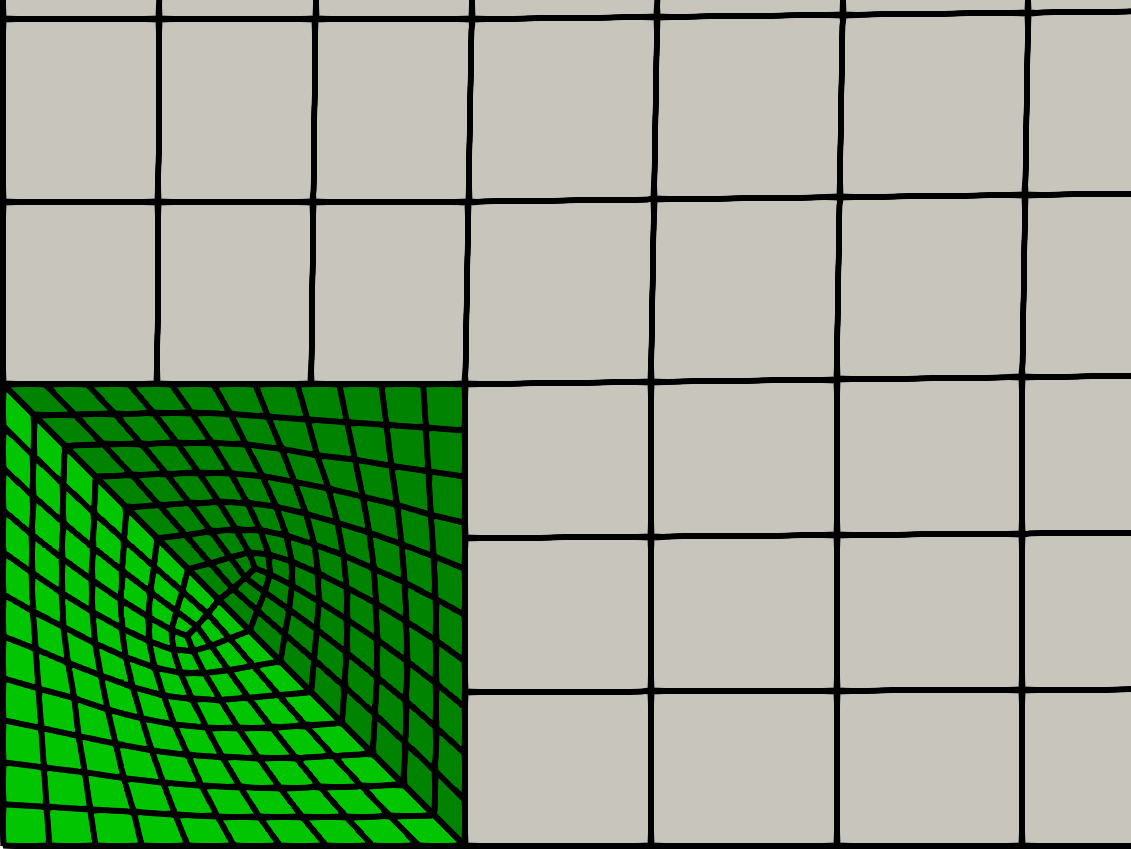}
        \caption{$\lambda/9$ mesh.}
        \label{fig:mesh-lambda9}
    \end{subfigure}
    \hspace{1cm}
    \begin{subfigure}[b]{0.25\textwidth}
        \centering
        \includegraphics[width=\textwidth,keepaspectratio,trim=0cm 0cm 0cm 0cm,clip]{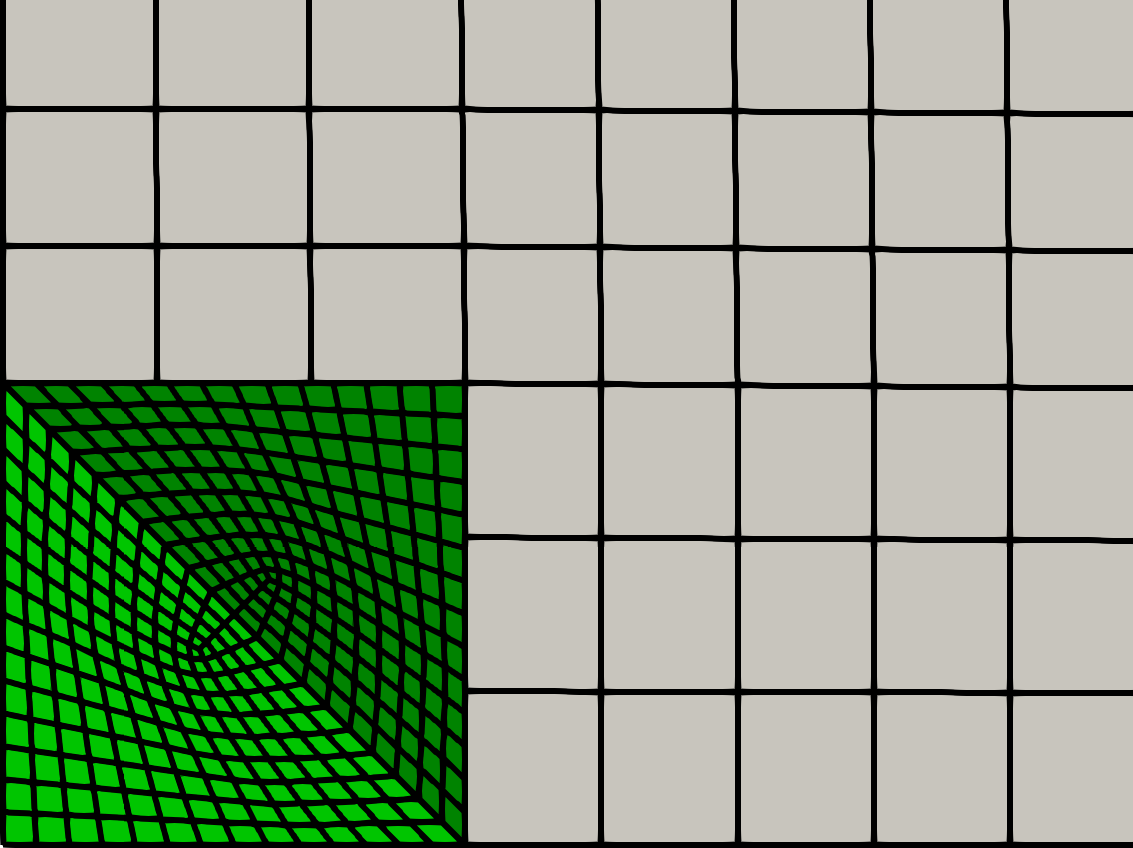}
        \caption{$\lambda/12$ mesh.}
        \label{fig:mesh-lambda12}
    \end{subfigure}
    \caption{Detail view of the cross-section in the $yz$-plane through the nonconforming meshes used for the convergence study. The absorber volumes $\Omega_{\mathrm{abs},1}$ and $\Omega_{\mathrm{abs},2}$ are colored greenish, and the air volume $\Omega_{\mathrm{air},1}$ is colored greyish.}
    \label{fig:convergence-meshes}
\end{figure}

For the convergence study, four different meshes have been investigated, whose element sizes correspond to the wavelength $\lambda$ at $f=\SI{200}{\hertz}$, i.e., the approximate element sizes were chosen as $\lambda / 3$, $\lambda / 6$, $\lambda / 9$, and $\lambda / 12$ in both $\Omega_\text{abs}$ and $\Omega_\text{air}$, as indicated in tab.~\ref{tab:convergence-study}. A crinkle clip view through the meshes is depicted in fig.~\ref{fig:convergence-meshes}. To quantify the discretization error, the relative frequency error measure $Err_{\mathrm{rel},f}^{L_2}$ is used, such that
\begin{equation}
    Err_{\mathrm{rel},f}^{L_2} = \sqrt{\frac{\sum_{i=1}^{N_\mathrm{modes}} \left( f_{\mathrm{mode,analyt}}^{(i)} - f_{\mathrm{mode,sim}}^{(i)} \right)^2}{\sum_{i=1}^{N_\mathrm{modes}} \left( f_{\mathrm{mode,analyt}}^{(i)}\right) ^2}} \, .
\end{equation}
Thereby, the first $N_\mathrm{modes}=277$ non-zero modal frequencies are considered (the 277th modal frequency being at \SI{199.43}{\hertz}), where $f_{\mathrm{mode,sim}}$ are the modal frequencies evaluated from a modal FE analysis, and $f_{\mathrm{mode,analyt}}$ are the analytically computed modal frequencies resulting from the ascending order of $f_{\mathrm{mode,analyt}}^{(n_x,n_y,n_z)}$ computed from eq.~\eqref{eq:analyt-modes}.

In tab.~\ref{tab:convergence-study}, the approximate element sizes, number of elements per subdomain, and the total number of elements of the different meshes are listed, as well as the respective computational cost (wall time and RAM\footnote{random-access memory}) evaluated on a computation server (running 24 CPU\footnote{central processing unit} threads on server "RK10" with two Intel (R) Xeon X5690 (6x 3,46 GHz each) CPUs) and the relative error measure $Err_{\mathrm{rel},f}^{L_2}$.

For decreasing element sizes, the mesh study exhibits a convergence towards the analytically calculated mode frequencies by means of a decreasing error measure, from $Err_{\mathrm{rel},f}^{L_2}=1.3245 \cdot 10^{-3}$ for the $\lambda / 3$-mesh to $Err_{\mathrm{rel},f}^{L_2}=4.4816\cdot 10^{-6}$ for the $\lambda / 12$-mesh, as listed in tab.~\ref{tab:convergence-study}. However, the increase in accuracy does not justify the extensive increase by a factor of approximately $64$ in wall time duration and a factor of $8$ in RAM demand comparing the $\lambda / 3$-mesh with the $\lambda / 12$-mesh. Balancing computational cost and accuracy, it is determined that the decrease in numerical error from the $\lambda / 6$-mesh to the $\lambda / 12$-mesh does not justify the increase in computational cost.
Therefore, it is concluded that results obtained with the $\lambda / 6$-mesh are sufficiently accurate for the simulation purpose.
\begin{table}[htbp]
    \centering
    \caption{Mesh properties for convergence study. The approximate wavelength $\lambda$ is evaluated at $f=\SI{200}{\hertz}$. The $\lambda / 6$-mesh is used for further evaluations.}
    \begin{tabular}{@{}lrrrrrrrrrr@{}}
    \toprule
        \multirow{2}{*}[-\aboverulesep]{mesh} &
        \multicolumn{3}{c}{approx. elem. size (m)} &
        \multicolumn{3}{c}{\# elements} &
        \multicolumn{1}{c}{\multirow{2}{*}[-\aboverulesep]{\shortstack[l]{elem.\\total}}} &
        \multicolumn{1}{c}{\multirow{2}{*}[-\aboverulesep]{\shortstack[l]{wall\\time}}} &
        \multicolumn{1}{c}{\multirow{2}{*}[-\aboverulesep]{\shortstack[l]{RAM\\(MB)}}} &  
        \multicolumn{1}{c}{\multirow{2}{*}[-\aboverulesep]{$Err_{\mathrm{rel},f}^{L_2}$}} \\
        \cmidrule(lr){2-4}\cmidrule(lr){5-7}
        & $\Omega_{\mathrm{air},1}$ & $\Omega_{\mathrm{abs},1}$ & $\Omega_{\mathrm{abs},2}$ & $\Omega_{\mathrm{air},1}$ & $\Omega_{\mathrm{abs},1}$ & $\Omega_{\mathrm{abs},2}$ & & & &  \\ 
         \midrule
        $\lambda / 3$ & 0.500   & 0.12 & 0.12 & $2\,023$ & $1\,050$ & $1\,050$ & $4\,123$ & 4m 47s & $2\,093$ & $\mathbf{1.3245 \cdot 10^{-3}}$  \\
        $\lambda / 6$ & 0.250   & 0.06 & 0.06 & $15\,708$ & $6\,255$ & $6\,255$ & $28\,210$ & 32m 37s & $3\,256$ & $\mathbf{8.6710 \cdot 10^{-5}}$  \\
        $\lambda / 9$ & 0.167  & 0.04 & 0.04 & $53\,550$ & $21\,109$ & $21\,109$ &  $95\,768$ & 2h 1m 26s & $7\,266$ & $\mathbf{1.5340\cdot 10^{-5}}$ \\
        $\lambda / 12$ & 0.125 & 0.03 & 0.03 & $124\,821$ & $50\,040$ & $50\,040$ & $224\,901$ & 5h 6m 53s & $16\,190$ & $\mathbf{4.4816\cdot 10^{-6}}$ \\
        \bottomrule
    \end{tabular}
    \label{tab:convergence-study}
\end{table}

\FloatBarrier
\subsection{Validation --- Point-Wise Comparison of Measurements and FE Simulations}

The simulation is validated by comparing TFs obtained from the simulation results to those obtained from measuring IRs for each of the four configurations depicted in fig.~\ref{fig:absorber-configs}. 

\begin{figure}[htbp]
    \centering
    \begin{subfigure}[c]{0.4\textwidth}
        \centering
        \includegraphics[width=\linewidth]{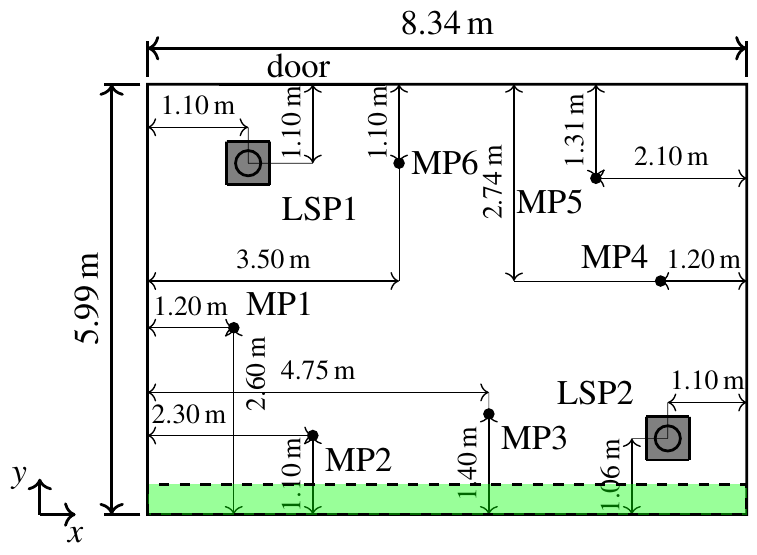}
        \caption{Floorplan coordinates of loudspeaker and microphone positions.}
        \label{fig:LSP-MP-2D}
    \end{subfigure}
    \hfill
    \begin{subfigure}[c]{0.4\textwidth}
        \centering
        \includegraphics[width=\linewidth]{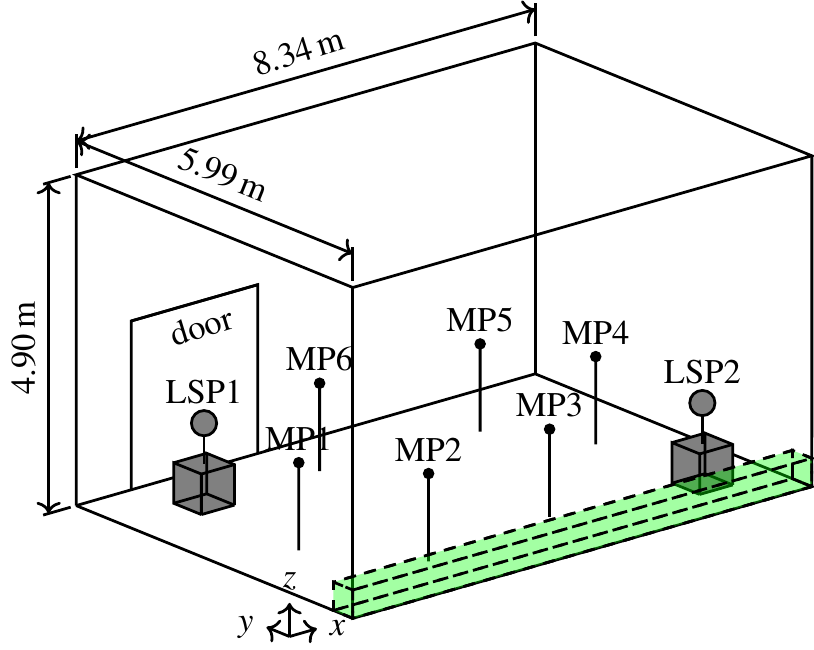}
        \caption{Schematic 3D-view of measurement setup.}
        \label{fig:LSP-MP-3D}
    \end{subfigure}
    \caption{Loudspeaker (LSP) and microphone (MP) positions for IR measurements \cite{Hofer2022Analyse}.}
    \label{fig:measurement-setup}
\end{figure}

\paragraph{Measurement Setup.}
The measurements were carried out as part of \cite{Hofer2022Analyse} and have been set up following reverberation time measurement requirements of ISO 354 \cite{2003ISO354} containing six microphone positions and two loudspeaker positions.
A class 1 certified 1/2'' NTi Audio M2230 omnidirectional condenser microphone is placed at a height of \SI{1.3}{\meter} at each microphone position (see fig.~\ref{fig:measurement-setup}). 
The excitation signal is played back below $f_c = \SI{120}{\hertz}$ via a Mackie SRM1850 subwoofer \cite{SRM1850} and above $f_c$ via a Norsonic NOR276 dodecahedron loudspeaker \cite{Nor276Nor280}.
During the measurement, both loudspeakers are located at the positions LSP1 and LSP2, respectively, as shown in fig.~\ref{fig:measurement-setup}, with the subwoofer on the floor and the dodecahedron loudspeaker at a height of \SI{1.3}{\meter} directly above the subwoofer.
The measurement microphones and the speakers are connected to a laptop (Lenovo ThinkPad T14s) via an RME Fireface UCX audio interface \cite{Carstens2018RME}. 
On the laptop, measurement control is performed using the ITA-Toolbox \cite{Berzborn2017ITA-Toolbox} in MATLAB. 

\begin{figure}[htbp]
    \centering
    \includegraphics[width=0.45\textwidth]{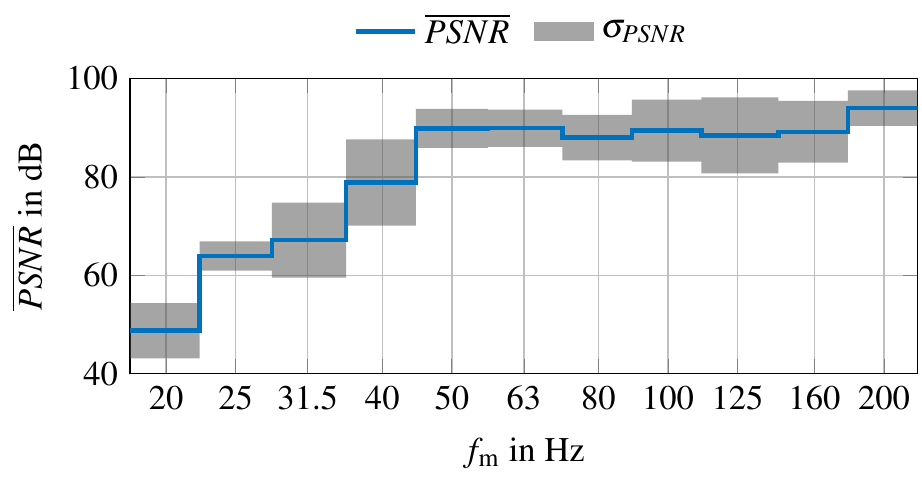}
    \caption{Third-octave band averaged Peak Signal-to-Noise-Ratio $\overline{PSNR}(f_m)$ with corresponding standard deviation $\sigma_{PSNR}$ from all IRs.}
    \label{fig:PSNR}
\end{figure}

The excitation signal is an exponential sinusoidal sweep with 2097152 samples played back in the frequency range from \SI{20}{\hertz} to \SI{24}{\kilo\hertz} at $f_s = \SI{48}{\kilo\hertz}$ resulting in a sweep length of $T = \SI{43.69}{\second}$.
The IR of each of the twelve transfer paths between the respective microphone and loudspeaker position is calculated by direct deconvolution of the recorded signal with the exponential sinusoidal sweep.
The \textit{Peak Signal-to-Noise-Ratio} $\overline{PSNR}(f_m)$, according to Lundeby \cite{Lundeby1995}, averaged over all IRs in the relevant frequency range ($\SI{20}{\hertz} \leq f_m \leq \SI{200}{\hertz}$) is shown in fig.~\ref{fig:PSNR}. 
Subsequently, the IRs are cropped to 30\,s and transformed in the frequency domain to obtain the twelve TFs with a frequency resolution of $\Delta f \approx \SI{0.03}{\hertz}$.
Since both the subwoofer and the dodecahedron loudspeaker do not have a linear amplitude frequency response (cf. \cite{SRM1850} and \cite{Nor276Nor280}), the obtained TFs are equalized with the inverse amplitude frequency responses of these in their corresponding playback frequency range.

\paragraph{FE Simulation Setup.}
The measurement setup was replicated in the harmonic FEM simulations by choosing $\vec{x}_\text{src}$ to be at the geometric center of the subwoofer loudspeaker at the loudspeaker positions LSP1 or LSP2 (see fig.~\ref{fig:LSP-MP-2D}), as the focus lies on the low frequency range. These simulations were performed on the same computer ("RK-10") as the grid study simulations, but were computationally more demanding: for example, using 22 CPU threads, EA1 with source position LSP1 had a wall time of 1h 10m and RAM usage of 3580 MB. The other configurations show very similar computational demand resulting in a total wall runtime of approximately 9.3h for all four EA configurations and two source positions.
The microphones are replicated by evaluation points (MP1 to MP6 in fig.~\ref{fig:LSP-MP-2D}), at which the acoustic pressure is interpolated using FE basis functions (a so-called sensor array). This enables a direct comparison between the TFs of simulations and measurements. The results of these evaluations are depicted in the appendix~\ref{sec:appendix-TF-plots}, where transfer functions obtained from (i) measurements and (ii) FE simulations of the four configurations, two LSP and six MP are depicted in figs.~\ref{fig:comp-TF-empty}, \ref{fig:comp-TF-config1}, \ref{fig:comp-TF-config2}, and \ref{fig:comp-TF-config3}.

\paragraph{Comparison of Measurements and Simulation Results.} Due to disparity of frequency resolution between TF measurements and simulations, further quantitative investigations are based on the \textit{third-octave band averaged spectra} $\bar{L}_{p,\mathrm{meas}}(f_\mathrm{m})$ of the measurements and $\bar{L}_{p,\mathrm{sim}}(f_\mathrm{m})$ of the simulations, where $f_\mathrm{m}$ is the third-octave band center frequency.

\begin{figure}[htbp]
	\begin{center}
	\scalebox{0.65}{\includegraphics{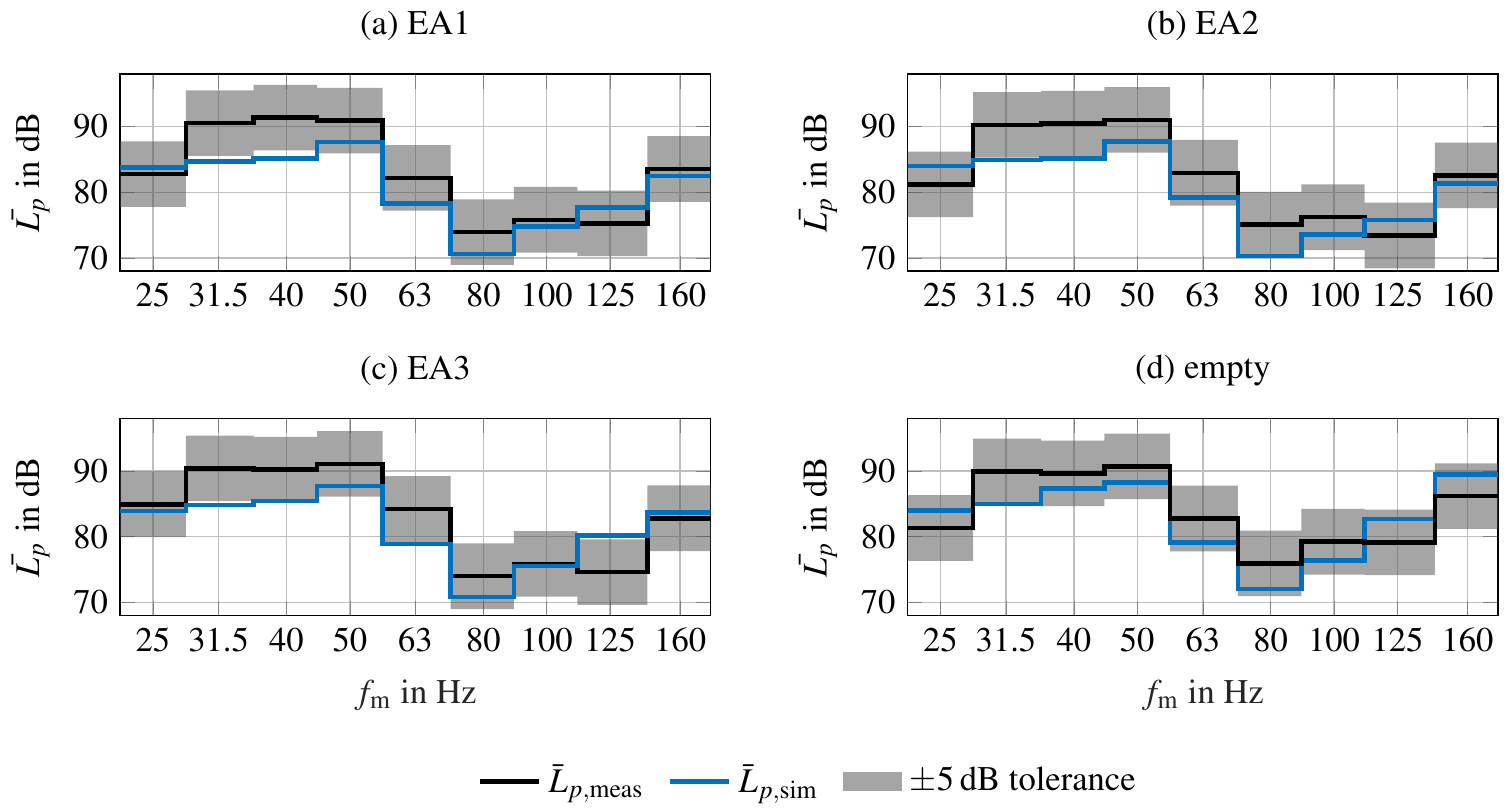}}
	\end{center}
   \caption{Third-octave band averaged TF $\bar{L}_{p,\mathrm{meas}}(f_\mathrm{m})$ of measurement and $\bar{L}_{p,\mathrm{sim}}(f_\mathrm{m})$ of FE simulation from LSP2 to MP2.}
   \label{fig:selected-thirdOct-TF}
\end{figure}

In fig.~\ref{fig:selected-thirdOct-TF}, the TF from LSP2 to MP2 is depicted exemplarily for all configurations. In all subplots, a good agreement between simulations and measurements is present, as the simulations lie in the $\pm\SI{5}{\decibel}$-interval around the measurements, for all third-octave band center frequencies except \SI{31.5}{\hertz} and \SI{40}{\hertz} in the configurations EA1, EA2, and EA3. For the empty RC depicted in the bottom right subplot of fig.~\ref{fig:selected-thirdOct-TF}, it is visible that in the third-octave bands above \SI{80}{\hertz}, the empty RC has a larger sound pressure level than the configurations EA1, EA2, or EA3. Some amplitude deviation between simulations and measurements is visible in the third-octave bands with the center frequencies $\SI{31.5}{\hertz}$ to $\SI{50}{\hertz}$: A reason for this (potentially systematic) deviation is presented below.

To further quantify the concordance between simulations and measurements, the error measure $Err_{L_p}$ is defined such that 
\begin{equation}
    Err_{L_p} = \frac{1}{N_{f_m}} \sum\limits_{f_\mathrm{m} \in \mathcal{F}}
    \left| \bar{L}_{p,\mathrm{meas}}(f_\mathrm{m}) - \bar{L}_{p,\mathrm{sim}}(f_\mathrm{m}) \right| \, ,
\end{equation}
where $\mathcal{F}$ is the set of the third-octave band center frequencies, and $N_{f_m}=9$ is the number of elements in $\mathcal{F}$. The error measure $Err_{L_p}$ is evaluated for each LSP/MP-combination resulting in twelve error values per EA configuration. Averaging across LSP/MP-combinations for each EA configuration results in the averaged error $\bar{E}rr_{L_p}$.

\begin{figure}[htbp]
	\begin{center}
    \scalebox{0.65}{\includegraphics{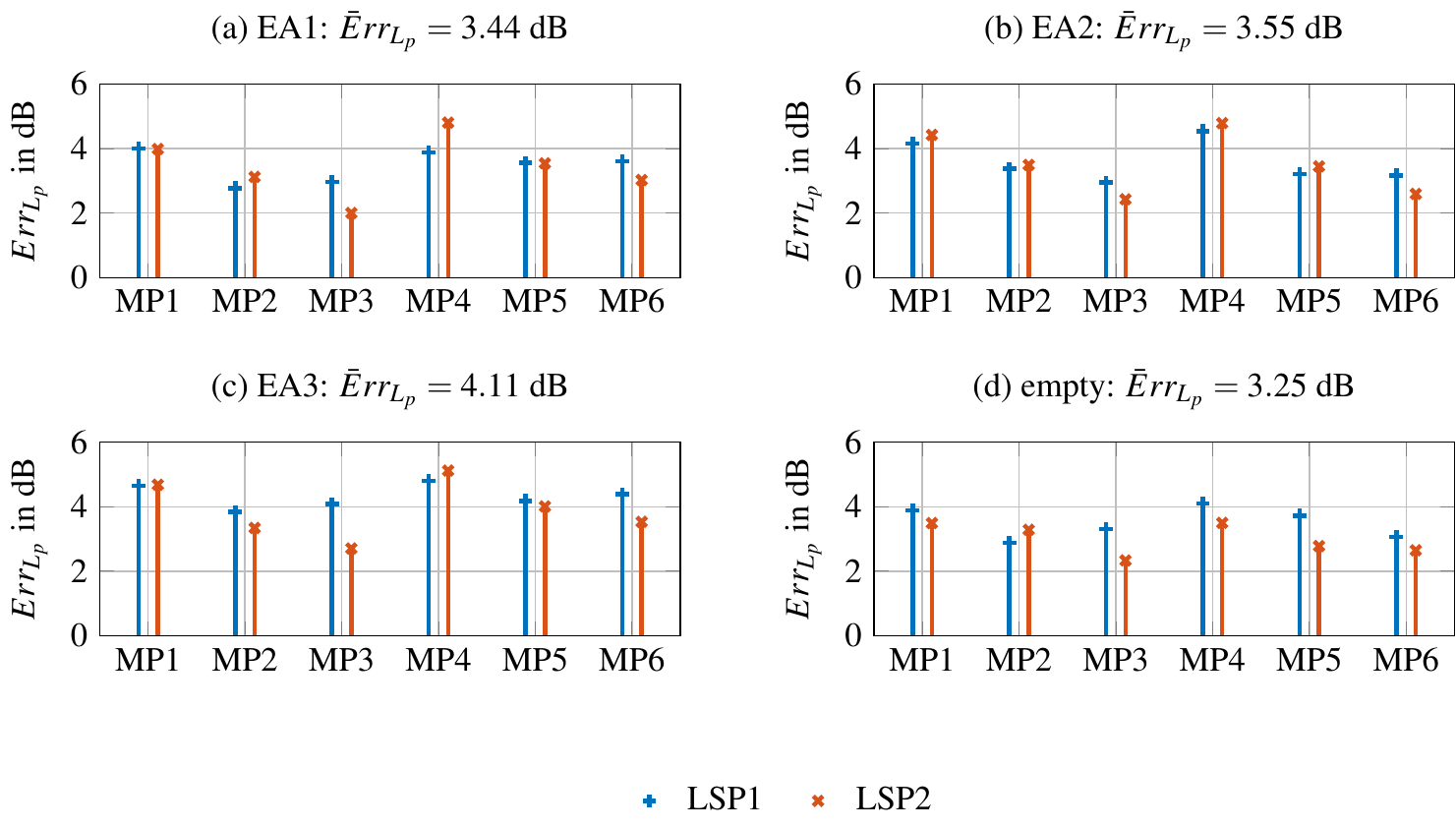}}
	\end{center}
   \caption{Third-octave band error $Err_{L_p}$ for all configurations and MP-LSP combinations. The averaged error measure $\bar{E}rr_{L_p}$ is obtained by averaging $Err_{L_p}$ across LSP/MP-combinations for each configuration.}
   \label{fig:error-measure}
\end{figure}

In fig.~\ref{fig:error-measure}, the error measure $Err_{L_p}$ is depicted for each LSP/MP-combination meaning that each stem in fig.~\ref{fig:error-measure} represents the third-octave band error $Err_{L_p}$ for \textit{one LSP/MP-combination of one configuration}. The top left subplot depicts $Err_{L_p}$ for configuration EA1, the top right subplot shows $Err_{L_p}$ for configuration EA2, the bottom left subplot shows $Err_{L_p}$ for configuration EA3, and the bottom right subplot shows $Err_{L_p}$ for the empty RC. Furthermore, the averaged error measure $\bar{E}rr_{L_p}$ is included for each configuration.
Comparing the error measure of the two LSP it can be concluded that no systematic error is present which can be traced back to a specific source position. From fig.~\ref{fig:error-measure} it is clear that the FE model achieves the lowest error for the empty RC (bottom right subplot), where the averaged error $\bar{E}rr_{L_p}$ is \SI{3.25}{\decibel}. For the different EA configurations, the averaged error increases in the following order: For EA1 the averaged error measure is $\bar{E}rr_{L_p}= \SI{3.44}{\decibel}$, for EA2 the averaged error measure is $\bar{E}rr_{L_p}=\SI{3.55}{\decibel}$, and finally for EA3 the averaged error measure is $\bar{E}rr_{L_p}=\SI{4.11}{\decibel}$. However, EA3 still exhibits a reasonable error value smaller than the $\SI{\pm 5}{\decibel}$-tolerance.
Therefrom it is evident, that the FE simulations of the empty room and all EA configurations are in good agreement with the measurements.

\begin{figure}[htbp]
    \centering
    \includegraphics[width=0.7\textwidth,keepaspectratio]{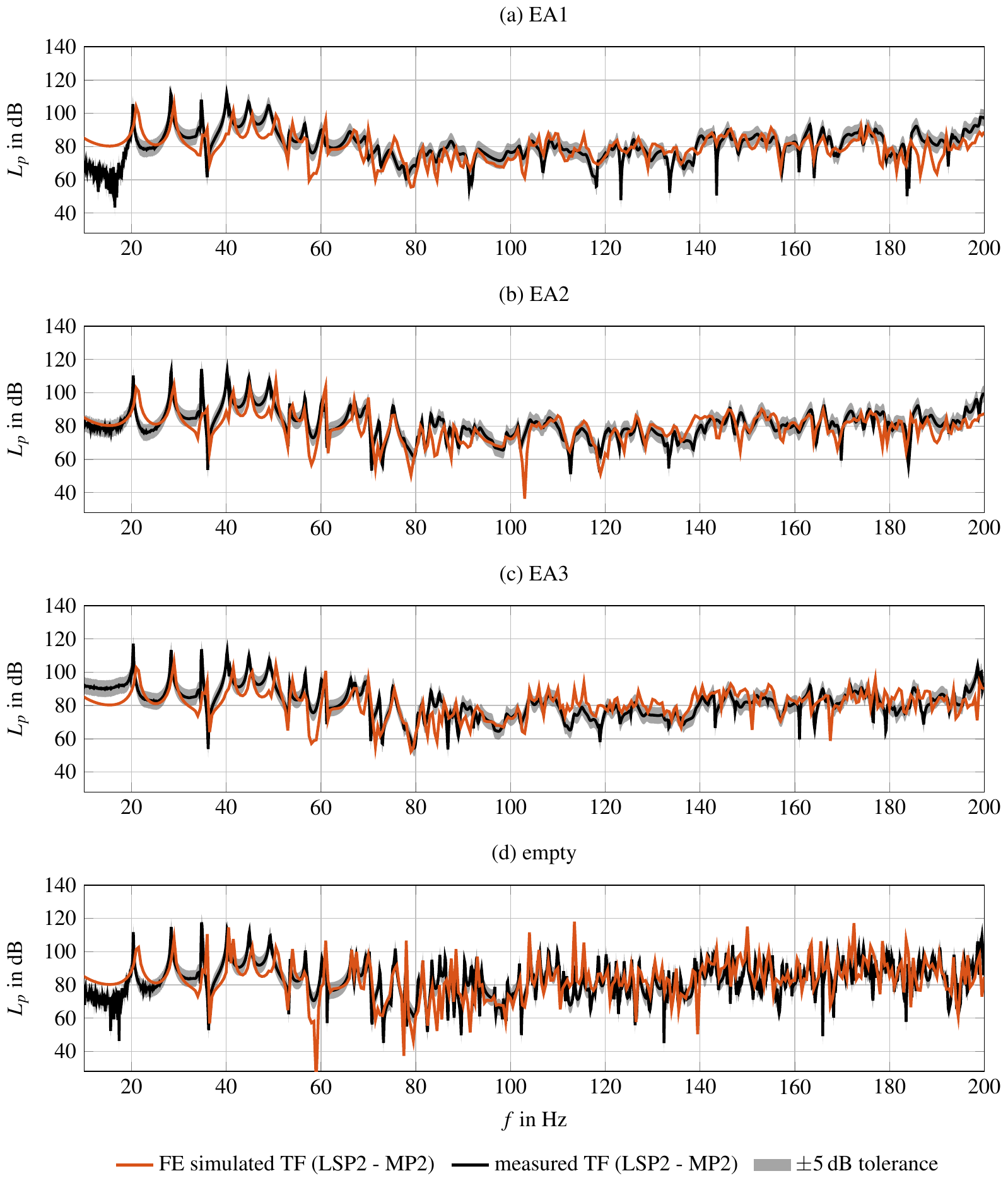}
    \caption{Comparison of measured and simulated TF between LSP2 and MP2 for all configurations.}
    \label{fig:comp_meas_sim_all_conf}
\end{figure}

The FE simulation allows for a detailed investigation of the resulting simulated TFs in comparison with the respective measured TFs, as visualized in fig.~\ref{fig:comp_meas_sim_all_conf}.
Figure~\ref{fig:comp_meas_sim_all_conf}~(a) shows the measured and simulated TFs from LSP2 to MP2 for configuration EA1. In fig.~\ref{fig:comp-TF-config1}, all other LSP/MP-combinations are visualized.
In fig.~\ref{fig:comp_meas_sim_all_conf}~(b), the measured and simulated TFs from LSP2 to MP2 for configuration EA2 are depicted. In fig.~\ref{fig:comp-TF-config2}, all remaining LSP/MP-combinations are visualized.
Figure~\ref{fig:comp_meas_sim_all_conf}~(c) depicts the measured and simulated TFs from LSP2 to MP2 for configuration EA3. In fig.~\ref{fig:comp-TF-config3}, the remaining LSP/MP-configurations are visualized.
In fig.~\ref{fig:comp_meas_sim_all_conf}~(d), the measured and simulated TFs from LSP2 to MP2 for the empty RC are depicted. The measured TF shows large fluctuations for very low frequencies (i.e., below \SI{20}{\hertz}) due to the measurement noise. All LSP/MP-combinations for the empty RC are depicted in fig.~\ref{fig:comp-TF-empty}.

When comparing fig.~\ref{fig:comp_meas_sim_all_conf}~(d) with fig.~\ref{fig:comp_meas_sim_all_conf}~(a) it is visible, that the edge absorber configuration EA1 damps frequencies above approximately \SI{60}{\hertz}, due to the modes in this frequency range being less pronounced in the configuration EA1 than in the empty room. A similar effect is visible when comparing configurations EA2 and EA3 in fig.~\ref{fig:comp_meas_sim_all_conf}~(b) and fig.~\ref{fig:comp_meas_sim_all_conf}~(c), respectively with the empty room in fig.~\ref{fig:comp_meas_sim_all_conf}~(d).
In all subplots of fig.~\ref{fig:comp_meas_sim_all_conf}, for the low frequency range (i.e., below approximately \SI{80}{\hertz} to \SI{100}{\hertz}), the room modes are clearly seperable. This is in line with the lower frequency bound of the RC, as defined in eq.~\eqref{eq:DeltaN}. Above this frequency, the number of modes per frequency interval is too high for individual modes being distinguishable.  
From the detailed comparison of measured and simulated TFs it can be concluded that some deviations between measurements and simulations are present, however the general tendencies and qualitative properties of the measurements are retained in the FE simulations with good agreement. Potential reasons for the deviations are discussed below.

For the sake of completeness, in appendix~\ref{sec:appendix-TF-plots}, the TFs of all individual evaluation positions are depicted for all configurations, which allow a detailed comparison of FE simulation results and measurements. In general, considering the intermediate conclusions above, it is visible, that the FE model is able to predict the modal characteristics of the sound field, especially for the low-frequency range of interest as well as the qualitative damping behaviour of the EA in the high-frequency range.

\paragraph{Reasons for Deviations Between Measurements and Simulations.}
In general, two types of errors can occur between measurements and simulations. Firstly, frequency errors, which are visible in the low frequency range (i.e., below \SI{80}{\hertz}) in all configurations, see figs.~\ref{fig:comp-TF-empty}, \ref{fig:comp-TF-config1}, \ref{fig:comp-TF-config2}, and \ref{fig:comp-TF-config3}. These become visible because the frequency resolution of the measurements is much higher than that of the FE simulations, and because of unmodeled effects such as the influence of the RC door. This problem is contained by introducing averaging over third-octave bands. Secondly, amplitude errors, which stem from many reasons, including measurement inaccuracies and unmodeled effects. To account for these errors, it is reasonable to introduce a tolerance band of, e.g., $\SI{\pm 5}{\decibel}$ in the third-octave averaged TFs depicted in fig.~\ref{fig:selected-thirdOct-TF}.

The unmodeled effects are as follows: The RC's walls, floor and ceiling are constructed from steel reinforced concrete, which is not perfectly sound-hard in the low-frequency regime (in the sense of a homogeneous Neumann boundary condition), as acoustic-structural interactions may be present. This was observed during the validation measurements as acoustic emissions emerged from the RC to the surrounding. Thus, the overall stiffness in the FE system is higher than in the real measurement chamber, and therefore the modal frequencies of the simulation are higher than the modal frequencies in the measured TFs. To verify these considerations by means of an additional mechanically-acoustically coupled FE simulation, material parameters (mass density and stress tensor) of the actual RC's walls must be known, but as they are not available, a verification simulation is not possible at the moment.

However, apart from the slight modal frequency deviations discussed above, the FE model is able to predict the acoustic TFs, between the loudspeaker and microphone positions used in the room acoustic measurements, for the empty room as well as for the three edge absorber configurations. 

\FloatBarrier
\section{Field Results}\label{sec:results}
\begin{figure}[htbp]
    \centering
    \begin{subfigure}[b]{0.3\textwidth}
        \centering
        \includegraphics[width=\textwidth,keepaspectratio,trim=0cm 0cm 0cm 0cm,clip]{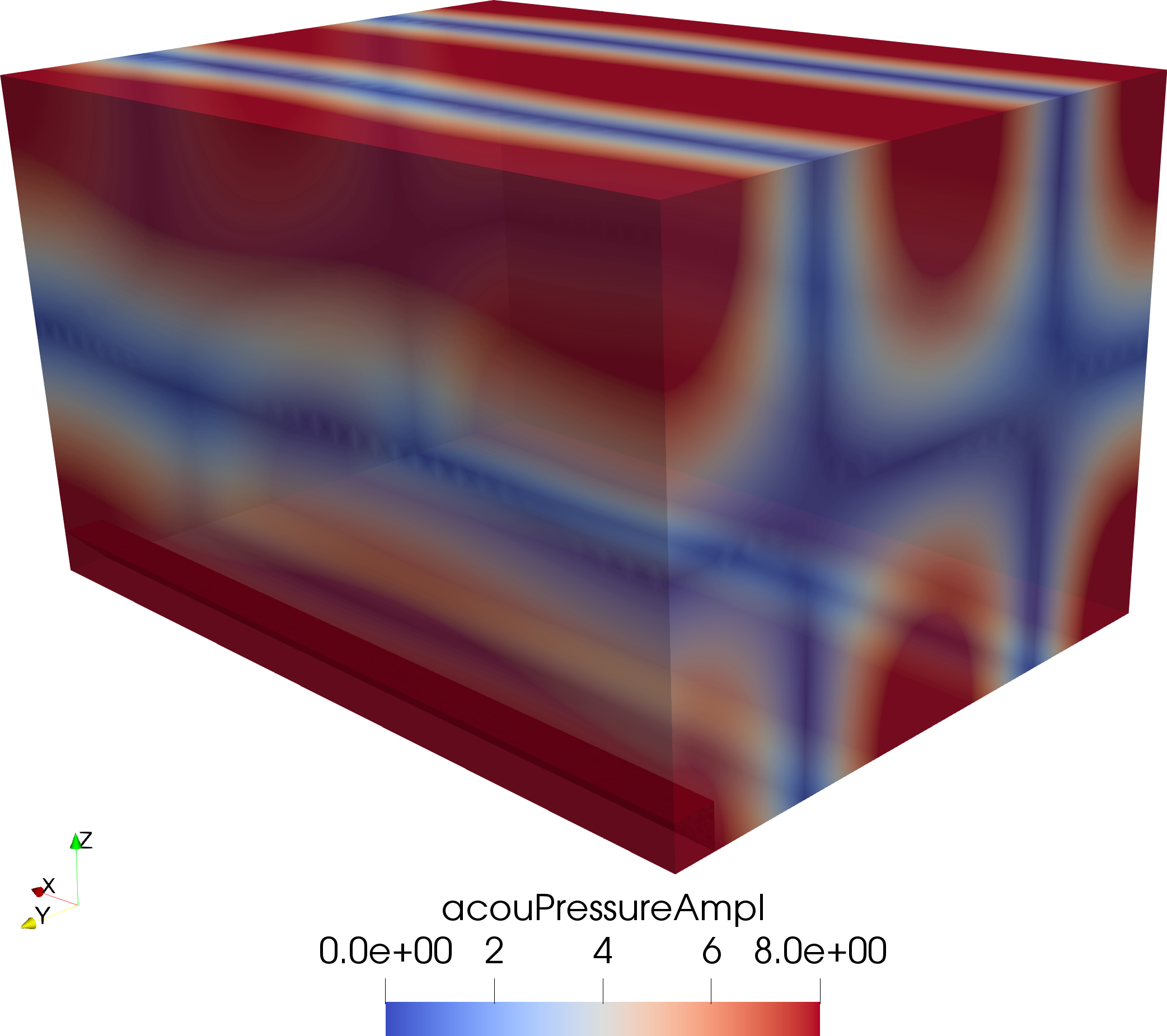}
        \caption{empty}
        \label{fig:}
    \end{subfigure}
    \hspace{1cm}
    \begin{subfigure}[b]{0.3\textwidth}
        \centering
        \includegraphics[width=\textwidth,keepaspectratio,trim=0cm 0cm 0cm 0cm,clip]{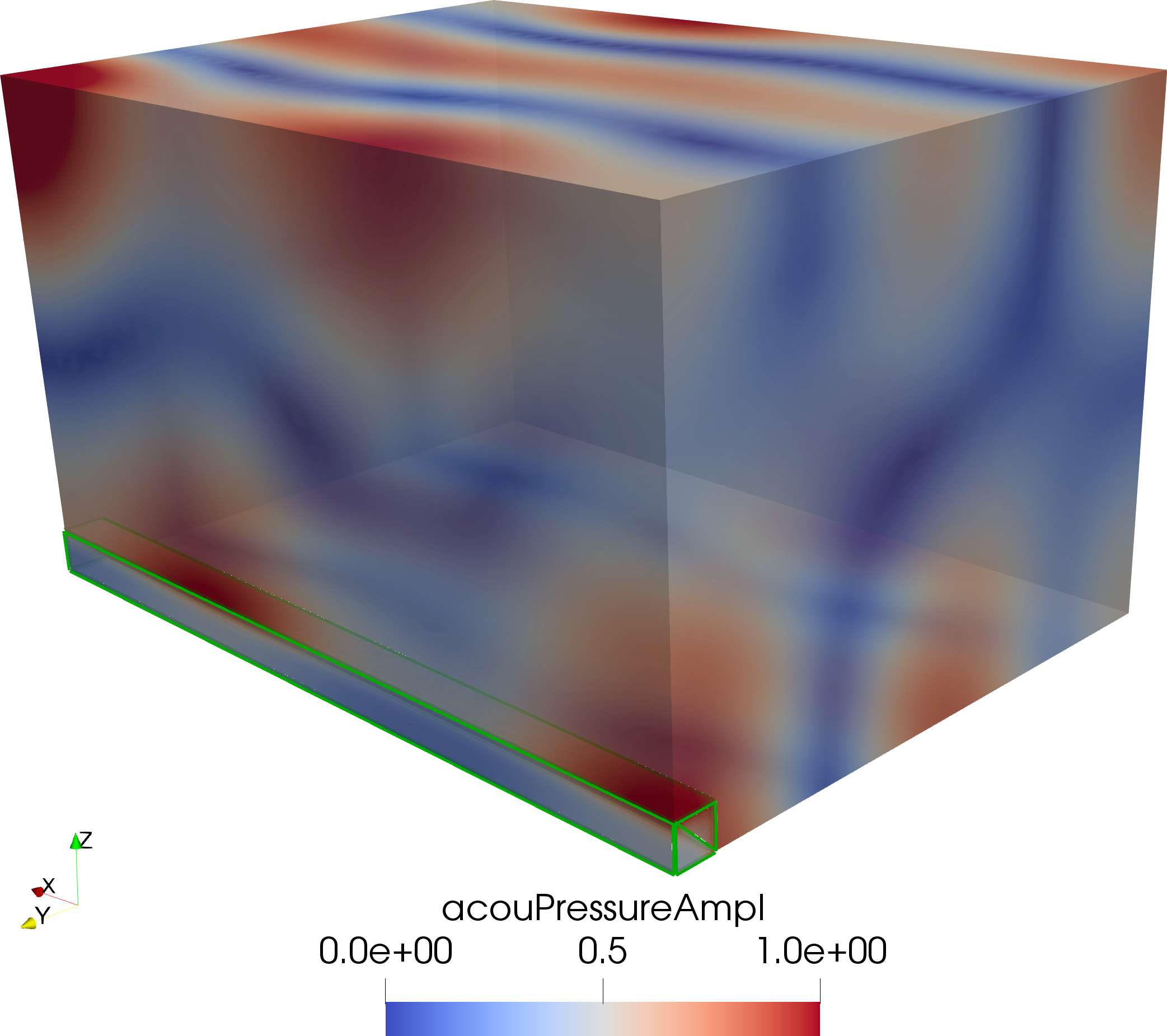}
        \caption{EA1}
        \label{fig:}
    \end{subfigure}\\
    \begin{subfigure}[b]{0.3\textwidth}
        \centering
        \includegraphics[width=\textwidth,keepaspectratio,trim=0cm 0cm 0cm 0cm,clip]{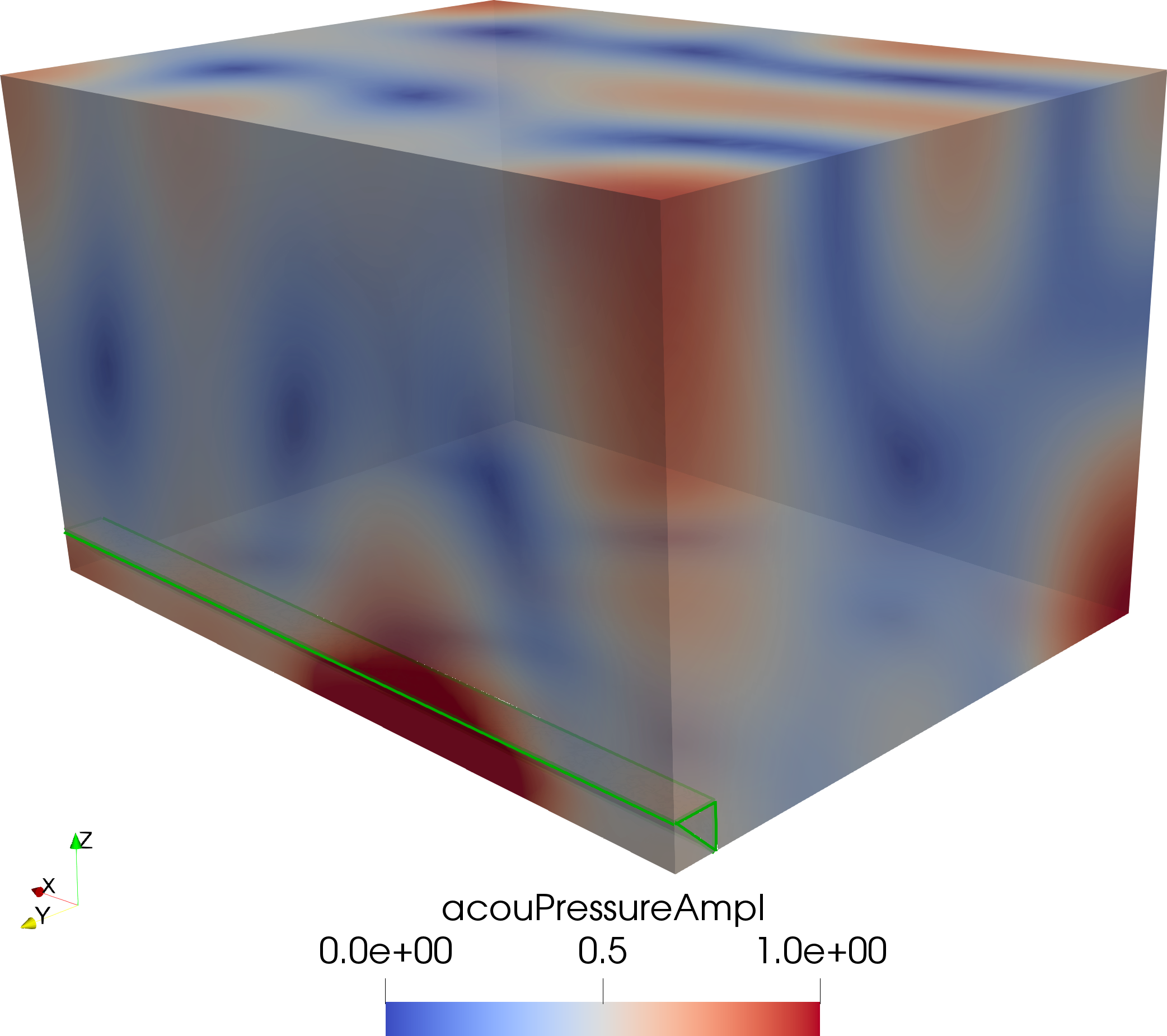}
        \caption{EA2}
        \label{fig:}
    \end{subfigure}
    \hspace{1cm}
    \begin{subfigure}[b]{0.3\textwidth}
        \centering
        \includegraphics[width=\textwidth,keepaspectratio,trim=0cm 0cm 0cm 0cm,clip]{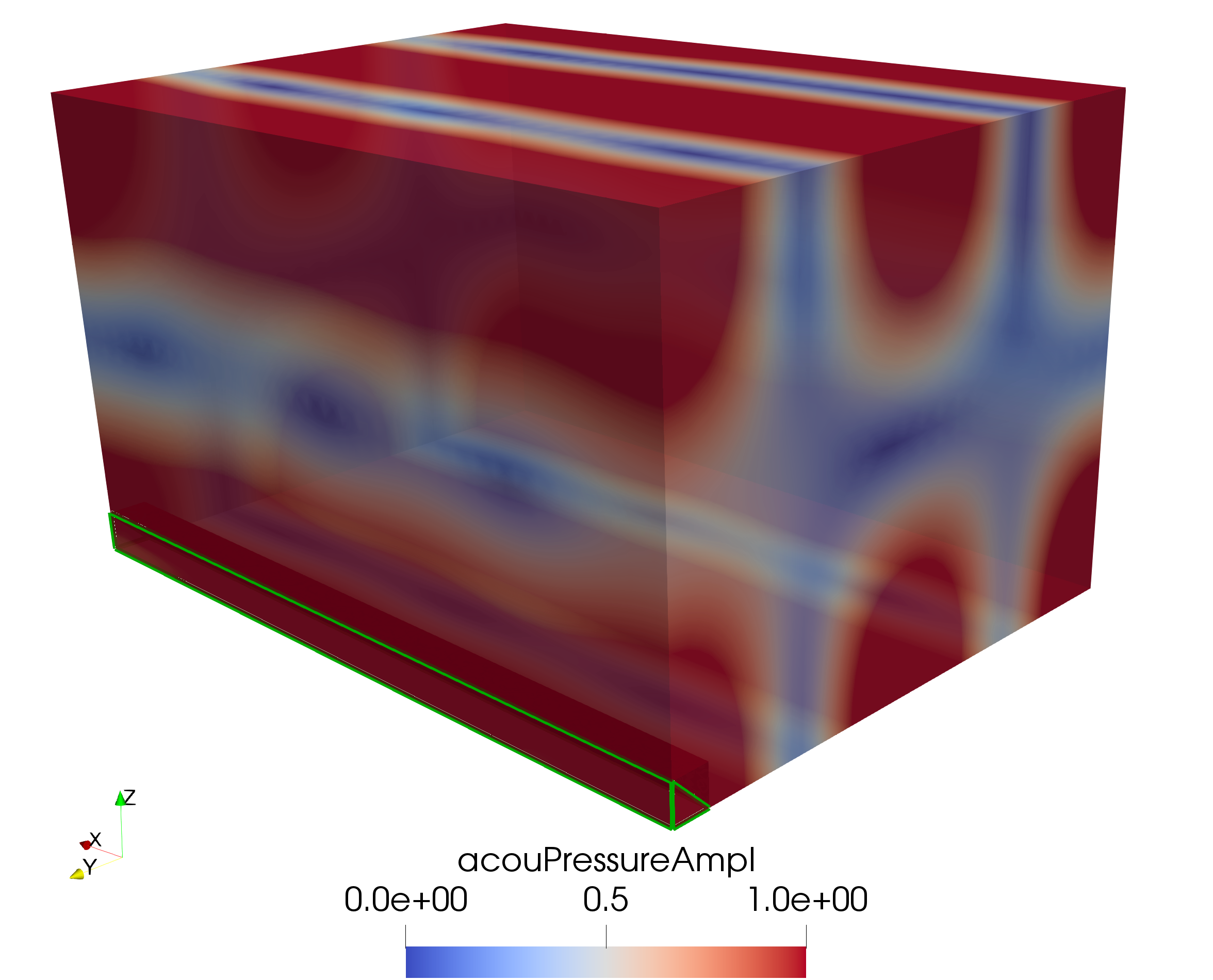}
        \caption{EA3}
        \label{fig:}
    \end{subfigure}
    \caption{Simulated acoustic pressure field at $f=\SI{66.5}{\hertz}$ excited at LSP1. The respective absorber volume $\Omega_\mathrm{abs}$ is colored greenish.}
    \label{fig:field-66.5}
\end{figure}

The validated FE model is used to obtain high resolution visualizations in the whole computational domain $\Omega$, including both air and absorber domains for all configurations depicted in fig.~\ref{fig:absorber-configs}. 
Figure~\ref{fig:field-66.5} shows the acoustic pressure field at $f=\SI{66.5}{\hertz}$ excited at LSP1 for all investigated configurations depicted in fig.~\ref{fig:absorber-configs}.

In fig.~\ref{fig:field-66.5}~(a), the pressure field in the empty room is depicted, which exhibits the expected modal shape of a room mode with the mode orders $(n_x, n_y, n_z)=(0,2,1)$ (see eq.~\eqref{eq:analyt-modes}). Looking at fig.~\ref{fig:field-66.5}~(b) and (c), the mode shape is distorted and the amplitude is damped, and the configuration EA2 exhibits qualitatively a better overall damping behavior as EA1. Figure~\ref{fig:field-66.5}~(d) depicts the resulting pressure field with configuration EA3, which is similar to the result of the empty RC (see fig.~\ref{fig:field-66.5}~(a)), but the amplitude is damped significantly. Overall, from fig.~\ref{fig:field-66.5}, it can be seen that all edge absorber configurations are able to damp the modal field amplitude at \SI{66.5}{\hertz} significantly, but EA1 and EA2 exhibit the best modal distortion capabilities.

\begin{figure}[htbp]
    \centering
    \begin{subfigure}[b]{0.3\textwidth}
        \centering
        \includegraphics[width=\textwidth,keepaspectratio,trim=0cm 0cm 0cm 0cm,clip]{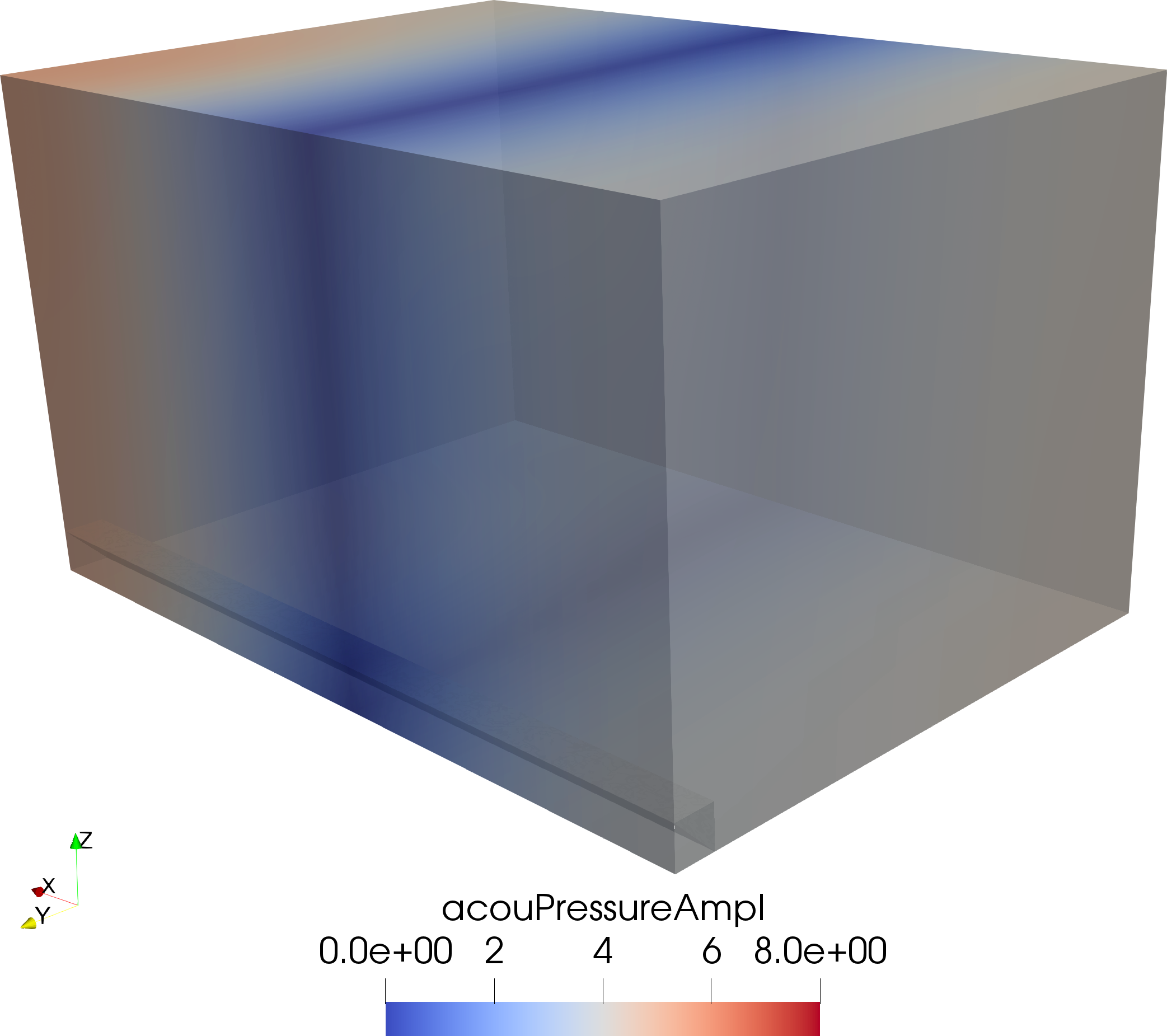}
        \caption{empty}
        \label{fig:}
    \end{subfigure}
    \hspace{1cm}
    \begin{subfigure}[b]{0.3\textwidth}
        \centering
        \includegraphics[width=\textwidth,keepaspectratio,trim=0cm 0cm 0cm 0cm,clip]{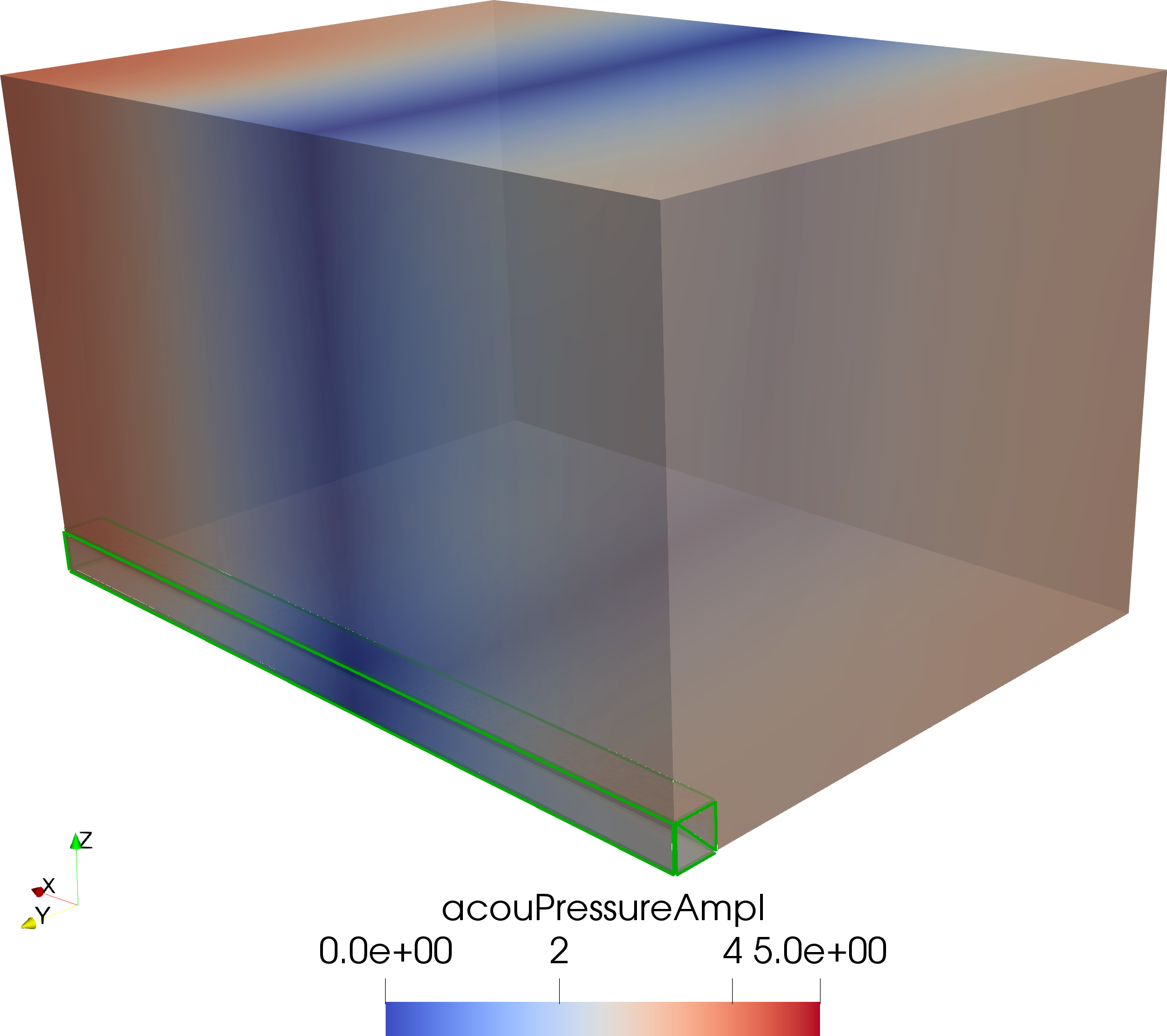}
        \caption{EA1}
        \label{fig:}
    \end{subfigure}\\
    \begin{subfigure}[b]{0.3\textwidth}
        \centering
        \includegraphics[width=\textwidth,keepaspectratio,trim=0cm 0cm 0cm 0cm,clip]{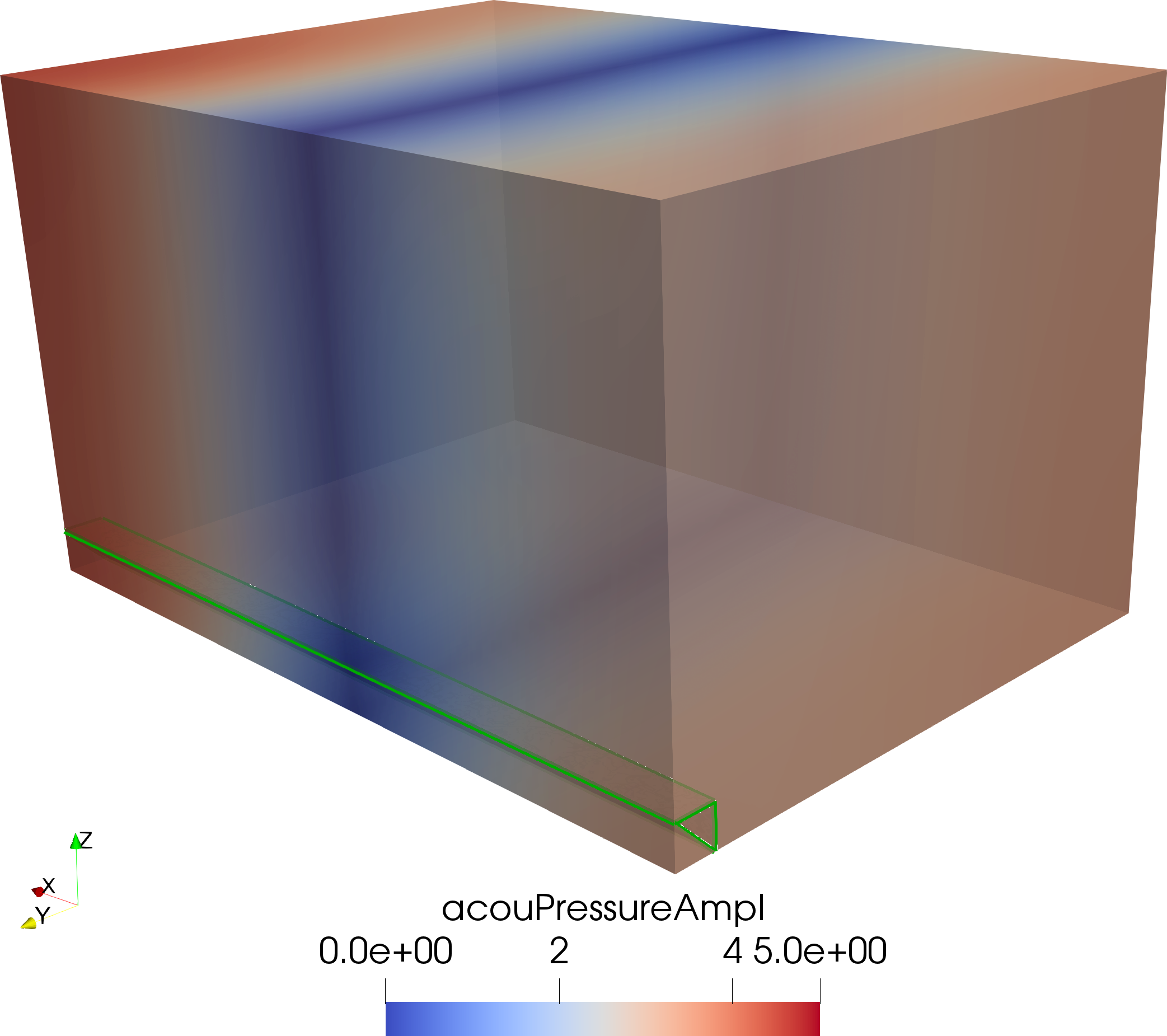}
        \caption{EA2}
        \label{fig:}
    \end{subfigure}
    \hspace{1cm}
    \begin{subfigure}[b]{0.3\textwidth}
        \centering
        \includegraphics[width=\textwidth,keepaspectratio,trim=0cm 0cm 0cm 0cm,clip]{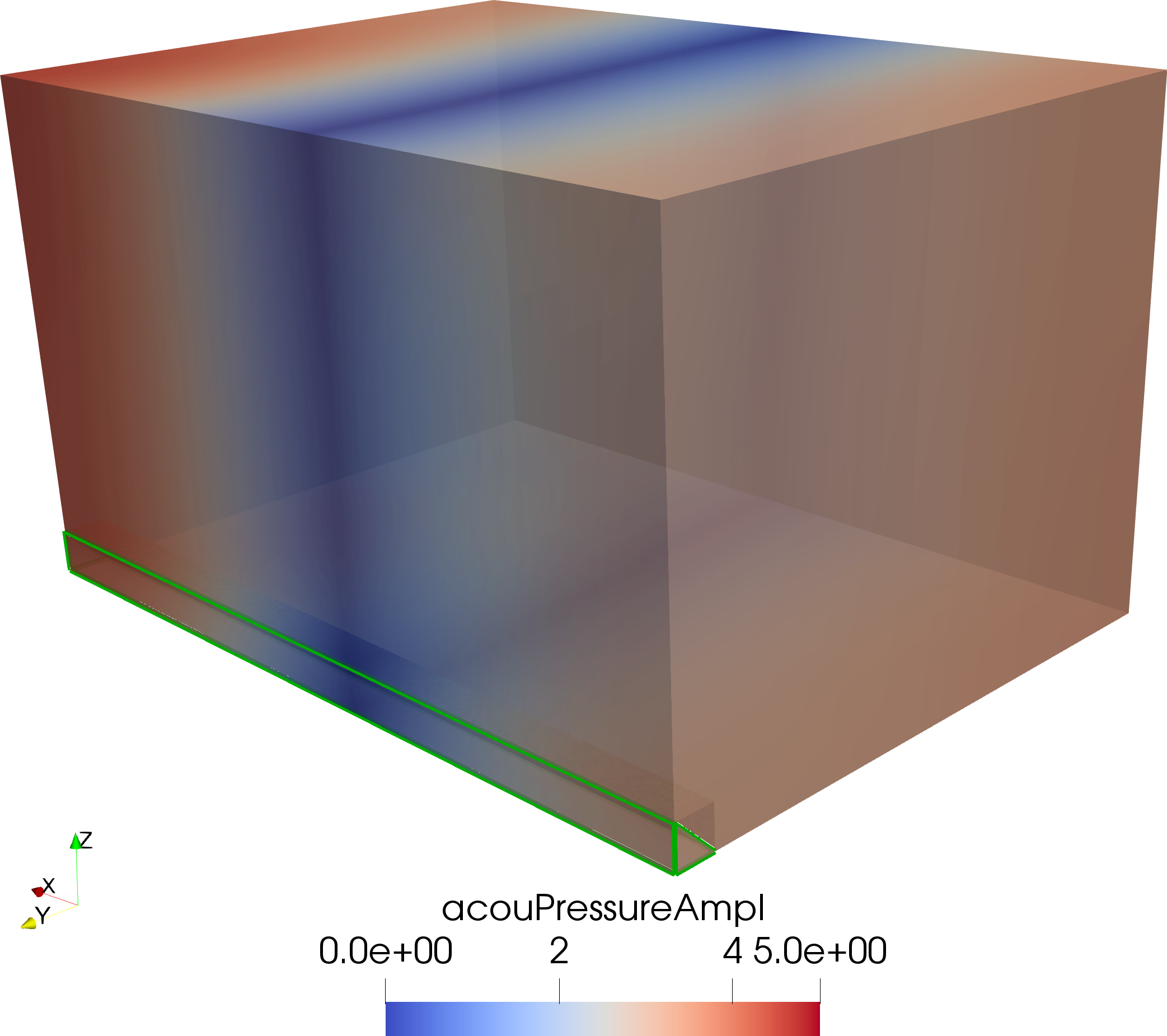}
        \caption{EA3}
        \label{fig:}
    \end{subfigure}
    \caption{Simulated acoustic pressure field at $f=\SI{21.5}{\hertz}$ excited at LSP1. The respective absorber volume $\Omega_\mathrm{abs}$ is colored greenish.}
    \label{fig:field-21.5}
\end{figure}

\begin{figure}[htbp]
    \centering
    \begin{subfigure}[b]{0.3\textwidth}
        \centering
        \includegraphics[width=\textwidth,keepaspectratio,trim=0cm 0cm 0cm 0cm,clip]{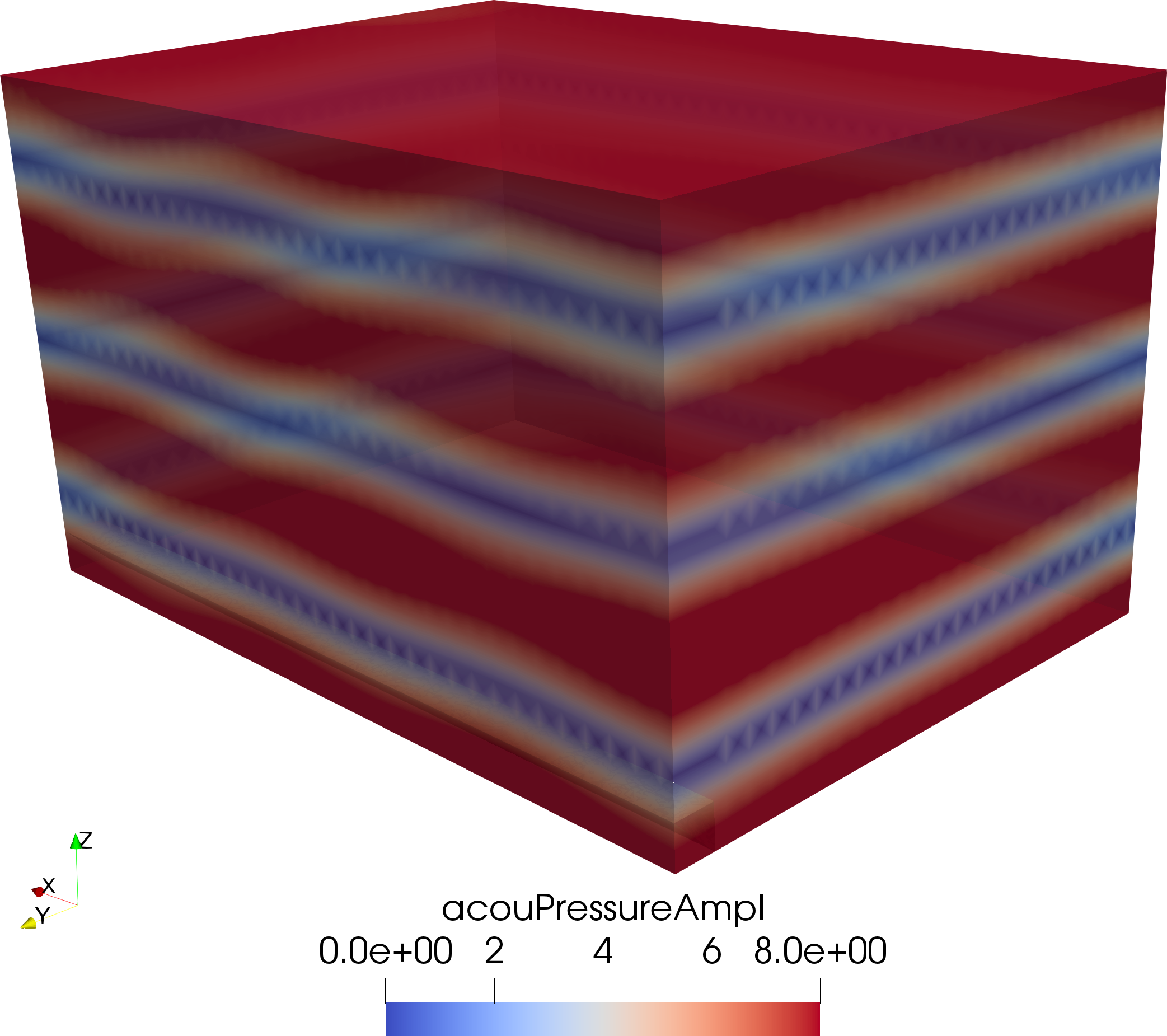}
        \caption{empty}
        \label{fig:}
    \end{subfigure}
    \hspace{1cm}
    \begin{subfigure}[b]{0.3\textwidth}
        \centering
        \includegraphics[width=\textwidth,keepaspectratio,trim=0cm 0cm 0cm 0cm,clip]{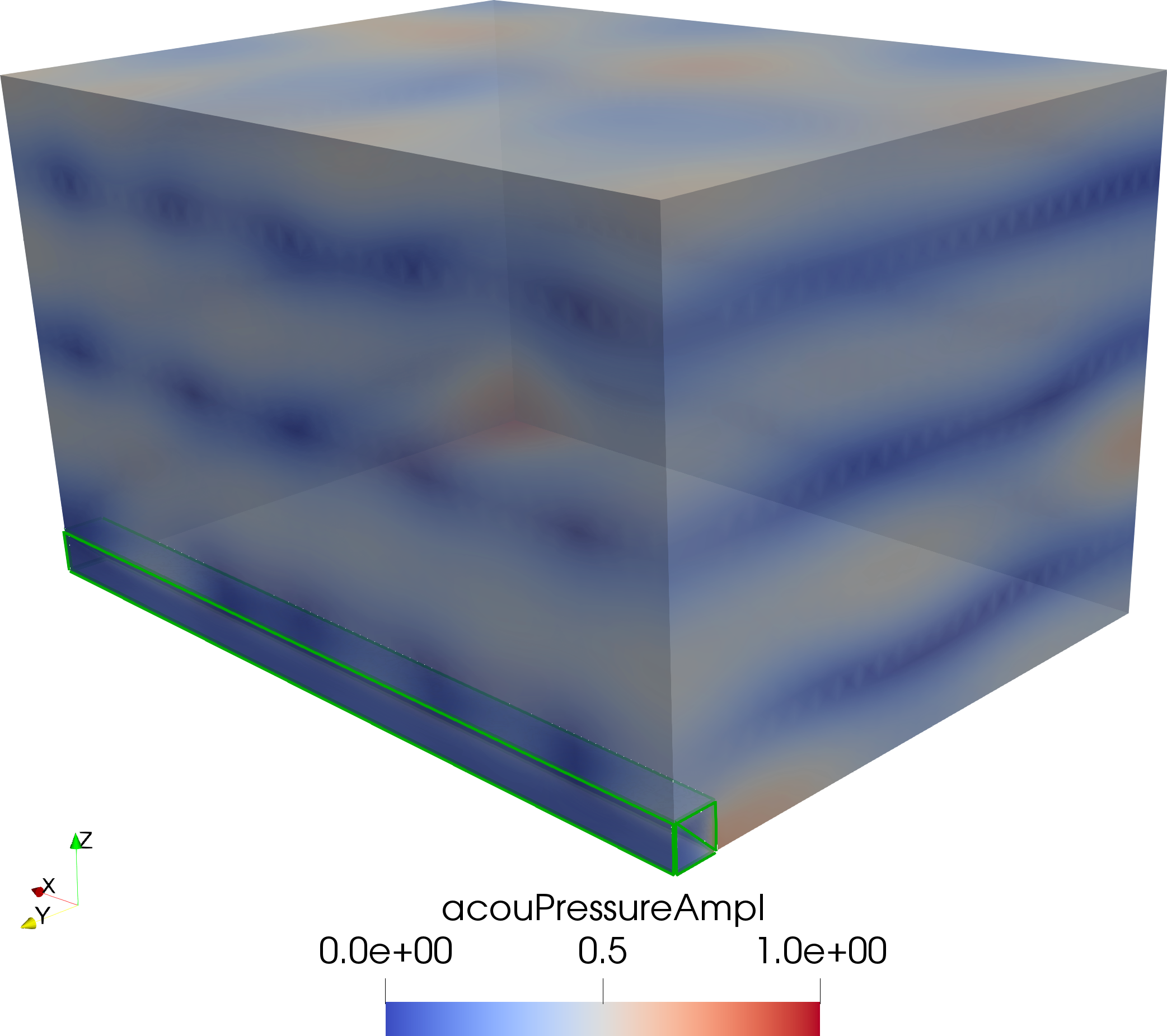}
        \caption{EA1}
        \label{fig:}
    \end{subfigure}\\
    \begin{subfigure}[b]{0.3\textwidth}
        \centering
        \includegraphics[width=\textwidth,keepaspectratio,trim=0cm 0cm 0cm 0cm,clip]{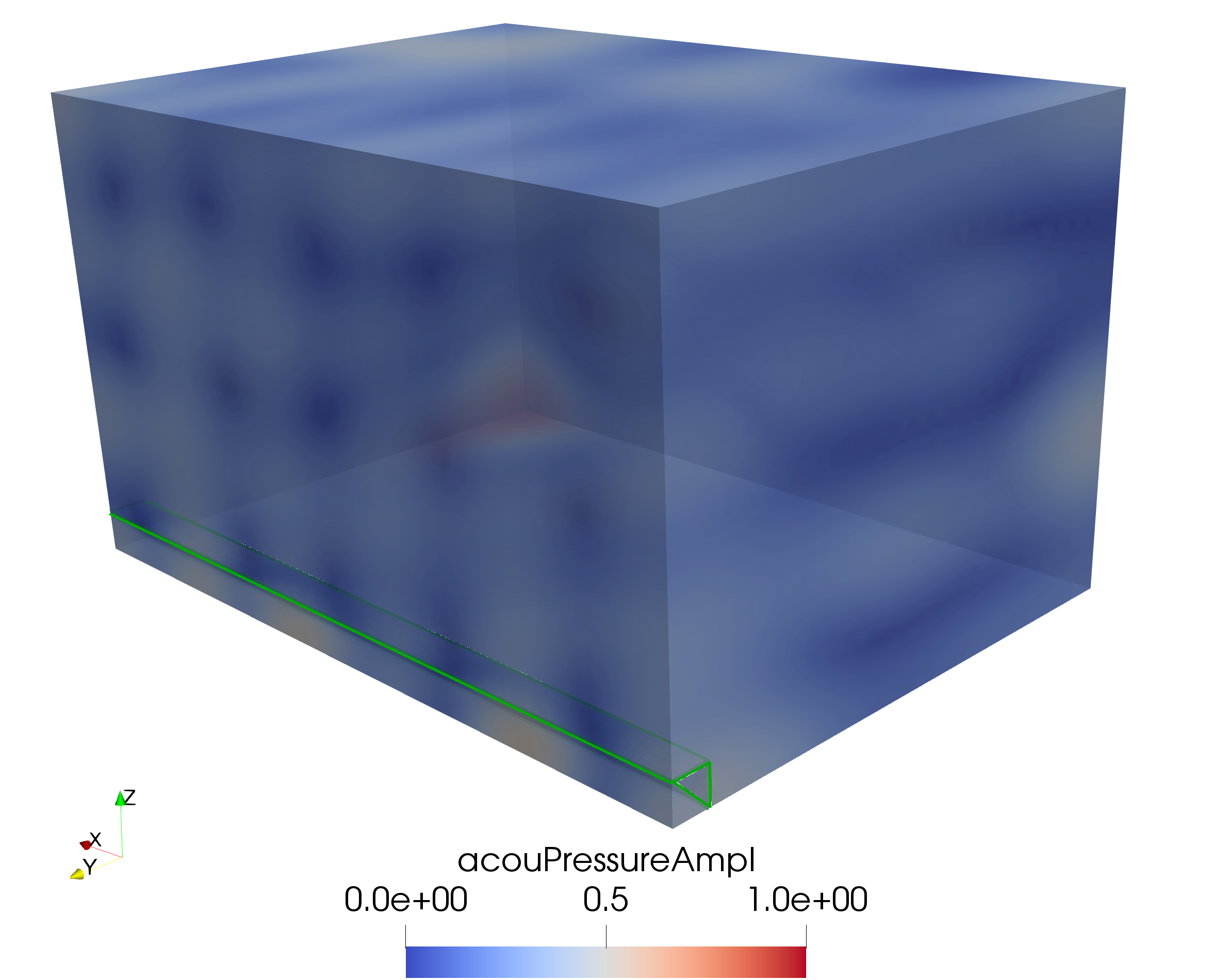}
        \caption{EA2}
        \label{fig:}
    \end{subfigure}
    \hspace{1cm}
    \begin{subfigure}[b]{0.3\textwidth}
        \centering
        \includegraphics[width=\textwidth,keepaspectratio,trim=0cm 0cm 0cm 0cm,clip]{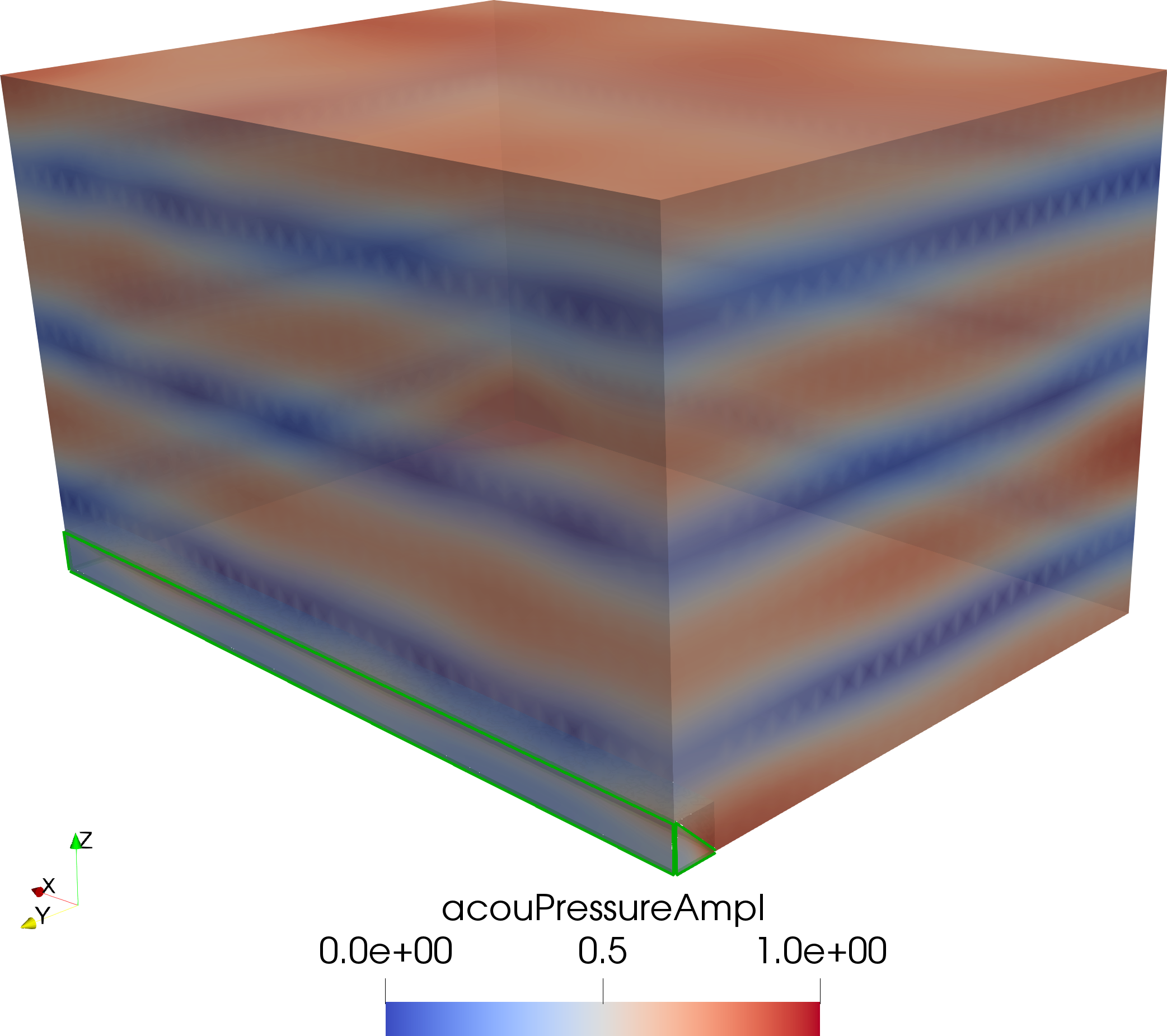}
        \caption{EA3}
        \label{fig:}
    \end{subfigure}
    \caption{Simulated acoustic pressure field at $f=\SI{104.0}{\hertz}$ excited at LSP1. The respective absorber volume $\Omega_\mathrm{abs}$ is colored greenish.}
    \label{fig:field-104.0}
\end{figure}

Additional field results are shown in fig.~\ref{fig:field-21.5} and \ref{fig:field-104.0}, in which the acoustic pressure field excited at LSP1 is depicted at $f=\SI{21.5}{\hertz}$ and $f=\SI{104.0}{\hertz}$, respectively, for all investigated configurations. In fig.~\ref{fig:field-21.5}, the pressure field of the $(1,0,0)$-mode is depicted. Comparing the empty room in fig.~\ref{fig:field-21.5}~(a) with configurations EA1, EA2, and EA3 shows that all EA configurations are able to damp the pressure amplitudes. Furthermore, comparing EA1 in fig.~\ref{fig:field-21.5}~(b) with both EA2 and EA3 in figs.~\ref{fig:field-21.5}~(c) and (d), it can be concluded that EA1 has a slightly better damping capability on the pressure amplitude than EA2 and EA3.
Figure~\ref{fig:field-104.0}~(a) shows the pressure field of the $(0,0,3)$-mode in the empty RC. Comparing EA3 in fig.~\ref{fig:field-104.0}~(d) with EA1 and EA2 (figs.~\ref{fig:field-104.0}~(b) and (c), respectively), it is visible that EA3 has a worse damping capability than EA1 and EA2.

\begin{figure}[htbp]
    \centering
    \begin{subfigure}[b]{0.4\textwidth}
        \centering
        \includegraphics[width=\textwidth,keepaspectratio,trim=0cm 0cm 0cm 0cm,clip]{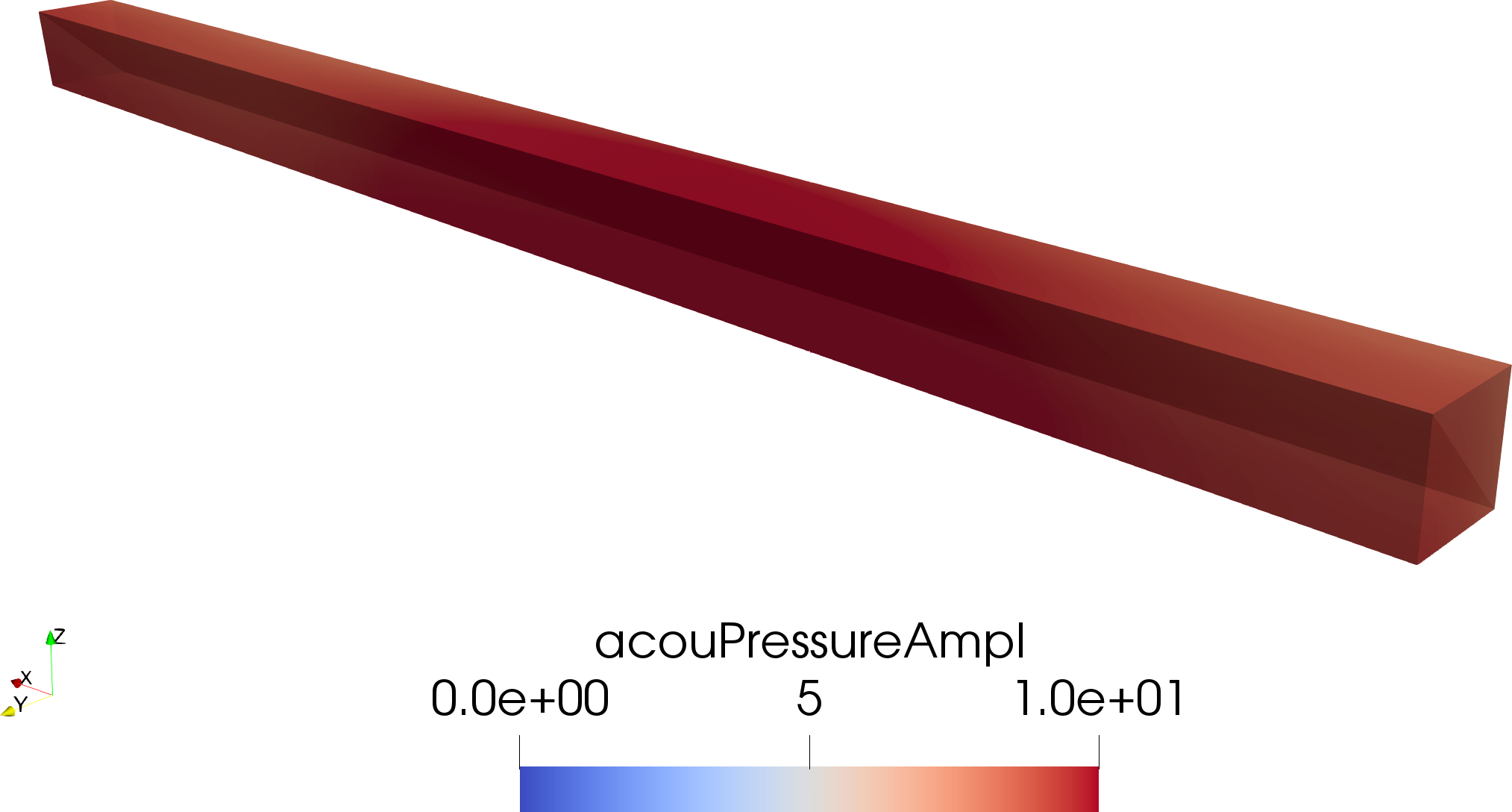}
        \caption{empty}
        \label{fig:}
    \end{subfigure}
    \hspace{1cm}
    \begin{subfigure}[b]{0.4\textwidth}
        \centering
        \includegraphics[width=\textwidth,keepaspectratio,trim=0cm 0cm 0cm 0cm,clip]{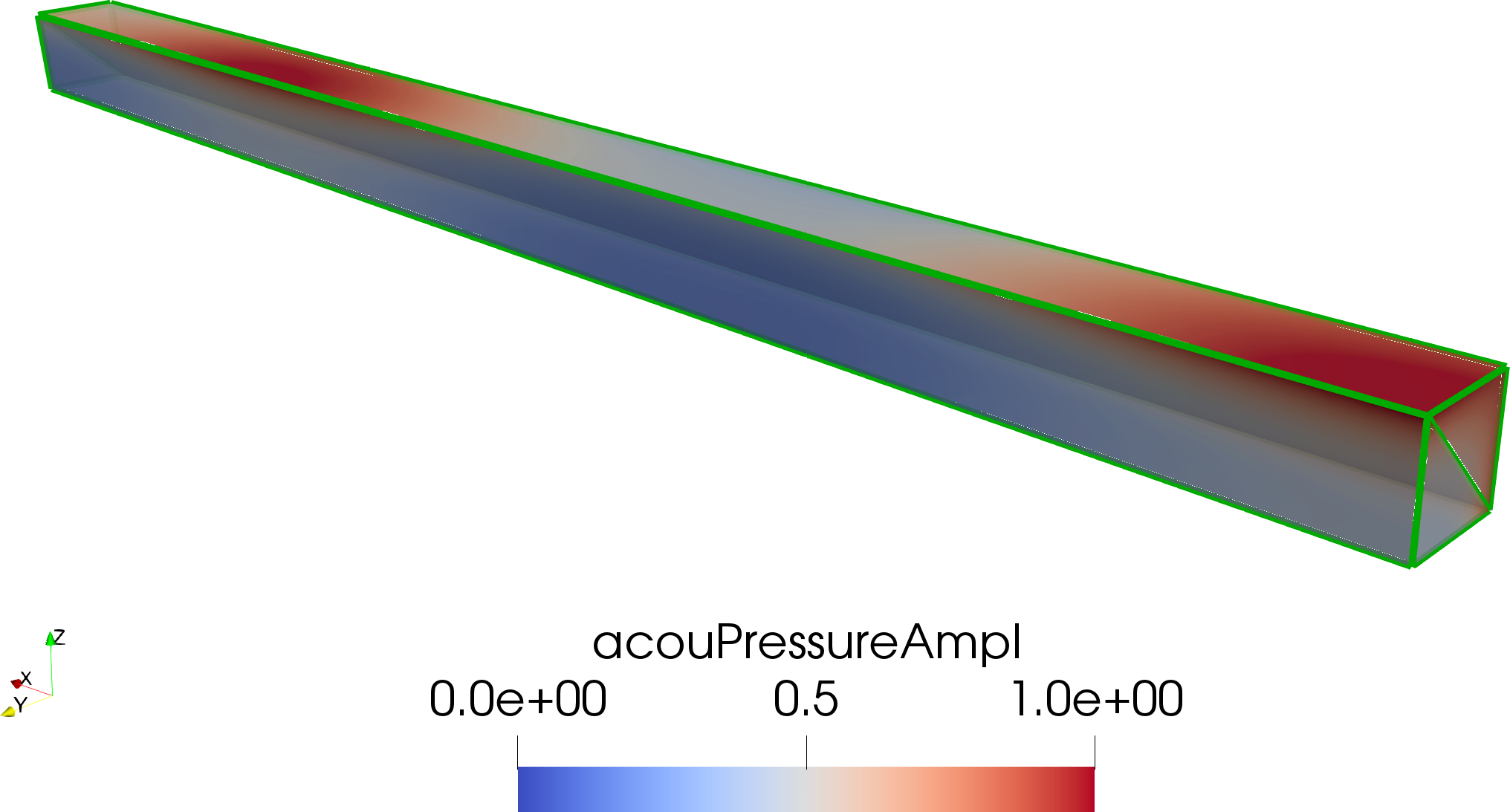}
        \caption{EA1}
        \label{fig:}
    \end{subfigure}\\
    \begin{subfigure}[b]{0.4\textwidth}
        \centering
        \includegraphics[width=\textwidth,keepaspectratio,trim=0cm 0cm 0cm 0cm,clip]{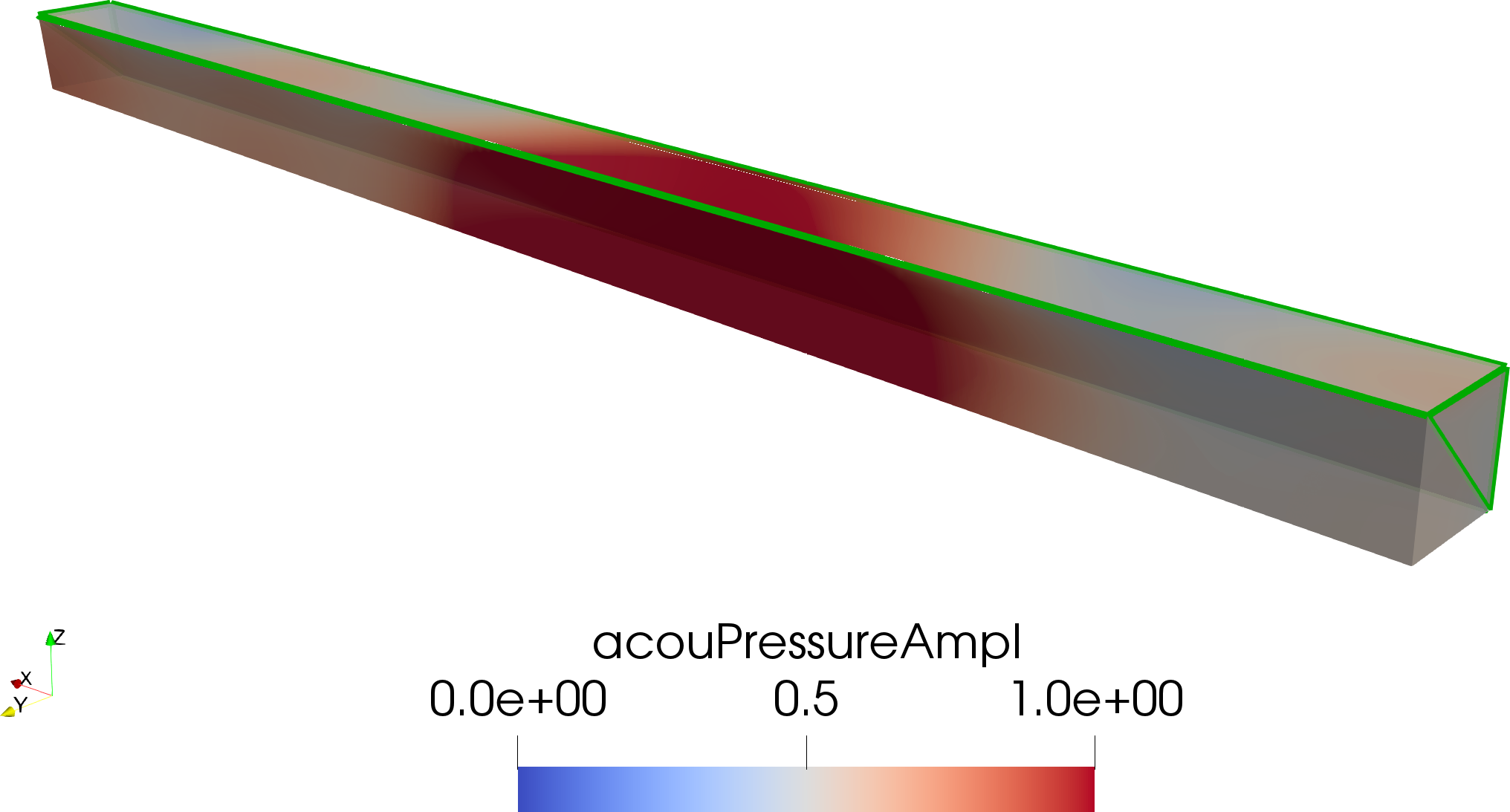}
        \caption{EA2}
        \label{fig:}
    \end{subfigure}
    \hspace{1cm}
    \begin{subfigure}[b]{0.4\textwidth}
        \centering
        \includegraphics[width=\textwidth,keepaspectratio,trim=0cm 0cm 0cm 0cm,clip]{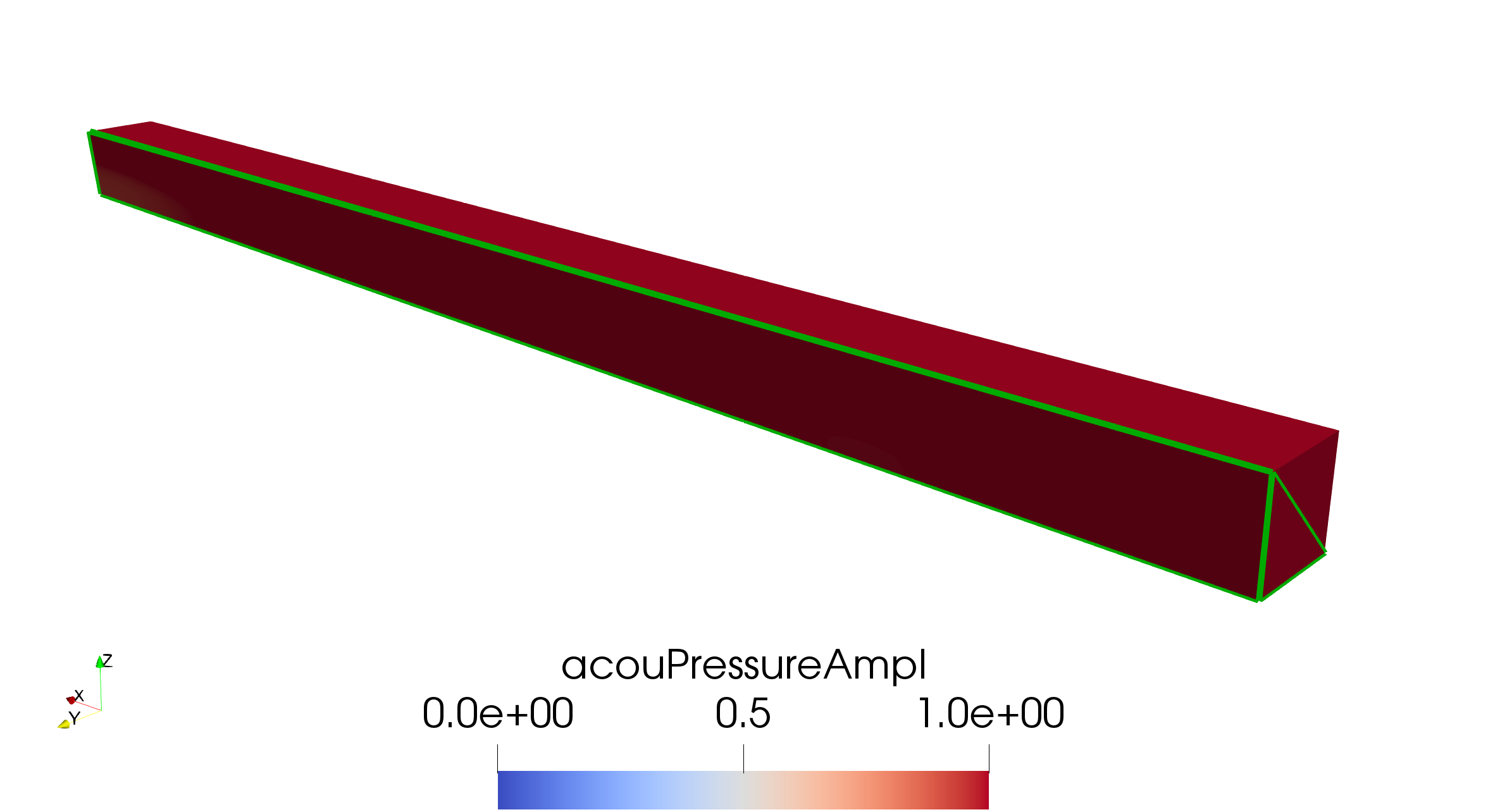}
        \caption{EA3}
        \label{fig:}
    \end{subfigure}
    \caption{Simulated acoustic pressure field in the volumes $\Omega_{\text{abs},1}$ and $\Omega_{\text{abs},2}$ at $f=\SI{66.5}{\hertz}$ excited at LSP1. The respective absorber volume $\Omega_\mathrm{abs}$ is colored greenish.}
    \label{fig:field-66.5-inEA}
\end{figure}

In fig.~\ref{fig:field-66.5-inEA}, the acoustic pressure field is visualized only for regions $\Omega_{\text{abs},1}$ and $\Omega_{\text{abs},2}$ as defined in fig.~\ref{fig:geometry-sketch}, regardless of the material. The volume filled with absorber material is marked greenish in the subfigures of fig.~\ref{fig:field-66.5-inEA}. For the empty RC depicted in fig.~\ref{fig:field-66.5-inEA}~(a), no damping of the pressure field is visible (this was already visible in fig.~\ref{fig:field-66.5}~(a)). Looking at configurations EA1 and EA2 depicted in fig.~\ref{fig:field-66.5-inEA}~(b) and (c), respectively, it can be observed that the acoustic pressure is damped in the volume parts filled virtually with absorber material. 
As visible in fig.~\ref{fig:field-66.5-inEA}~(c), the air-filled backing volume allows for a better wave propagation (i.e., higher acoustic pressure) than the absorber material, which is concluded from the higher acoustic pressure in the backing volume than in the absorber. Looking at fig.~\ref{fig:field-66.5-inEA}~(d) is can be observed that in the absorber volume, the acoustic pressure is damped. But since very close to the edge a minimum of acoustic velocity is present \cite{Waterhouse1955Interference}, it is expected that the porous material acting on the acoustic particle velocity is less effective than with (i) a larger cross-section (i.e., EA1) or (ii) a placement which is not as close to the edge (i.e., EA2) is to be favored above configuration EA3.

To summarize, the FE result exhibits the typically reported behaviour of edge absorbers: while the empty room clearly has a pronounced modal field, in configurations EA1 and EA2, the modal field is dampened significantly. Less damping is visible for configuration EA3. This is in line with the findings reported in \cite{Kurz2020Edge}, where the edge absorber is described as a damping element onto the room modes (i.e., a so-called "modal brake").

\FloatBarrier
\section{Conclusion}\label{sec:conclusion}
The presented FE simulation model for room acoustic edge absorbers using the JCAL equivalent fluid model for porous material is able to predict the sound field at the validation measurement locations.
The convergence of the FE simulation model has been verified by comparing analytical eigenfrequencies with the results of an eigenfrequency analysis across a grid study for an empty room. Thereby, it has been shown that a non-conforming second order grid with a mesh size of $\lambda/6$ at approximately $f=\SI{200}{\hertz}$ achieves an error measure of $Err_{\mathrm{rel},f}^{L_2}=8.6710 \cdot 10^{-5}$, comparing analytically and numerically obtained eigenfrequency values. It has been shown that the frequency error measure $Err_{\mathrm{rel},f}^{L_2}$ decreases with decreasing mesh size, from which it is concluded that the FE model converges to the analytical solution.

The FE simulation model has been validated against TFs obtained from IR measurements in the RC of the Laboratory for Building Physics at Graz University of Technology. Investigating third-octave band averaged spectra, an averaged spectral error measure $\bar{E}rr_{L_p} = \SI{3.25}{\decibel}$ is achieved for the empty configuration, from which it is concluded that the empty RC can be modeled accurately by means of the FE simulation. Furthermore, the edge absorber configurations EA1, EA2, and EA3 are modeled with averaged spectral error measures of  $\bar{E}rr_{L_p} = \left\{ \SI{3.44}{\decibel}, \SI{3.55}{\decibel}, \SI{4.11}{\decibel}\right\}$, respectively. Therefrom it is concluded, that the FE model is able to describe the pressure field in the real RC both with and without EA with a high degree of accuracy. Furthermore, an explanation of deviations between measurements and FE results with respect to small modal frequency deviations is provided.

The field results exhibit the typical behavior of edge absorbers which is a damping and distortion of the modal pressure field of the empty RC. This supports the empirical observations of Fuchs and Lamprecht \cite{Fuchs2013Covered} as well as Kurz et al. \cite{Kurz2020Edge}. In addition, the validated FE model allows for visualizing the acoustic field in the absorber geometry, which is hard to achieve with measurements due to the non-negligible influence of the measurement equipment on the acoustic field.

The validated FE model may be used to achieve more insights and understanding of the interactions between the porous edge absorber and the sound field of the RC by investigating the sound field consisting of acoustic pressure and particle velocity inside the absorber volume. These investigations can be supported by additional measurements of the pressure field in close proximity to the absorber in order to achieve a precise understanding of physical consequences of using edge absorbers. In addition to that, an optimal edge absorber configuration with respect to absorber placement or required material may be achieved by topology optimization.

\section*{Acknowledgements}
The authors thank Robert Hofer and Leon Merkel for carrying out measurements and simulations, respectively, and the team of the Laboratory for Building Physics at TU Graz for access to and support with the RC. F.~K. received funding from the Austrian Research Promotion Agency (FFG) under the Bridge project No. 39480417.

\section*{Conflict of Interests}
The authors declared no conflicts of interests.

\FloatBarrier
\clearpage
\appendix
\section{Point-Wise Comparison of FE-Simulation and Measurements}
\label{sec:appendix-TF-plots}
\begin{figure}[htbp]
    \centering
    \includegraphics[width=\textwidth,keepaspectratio]{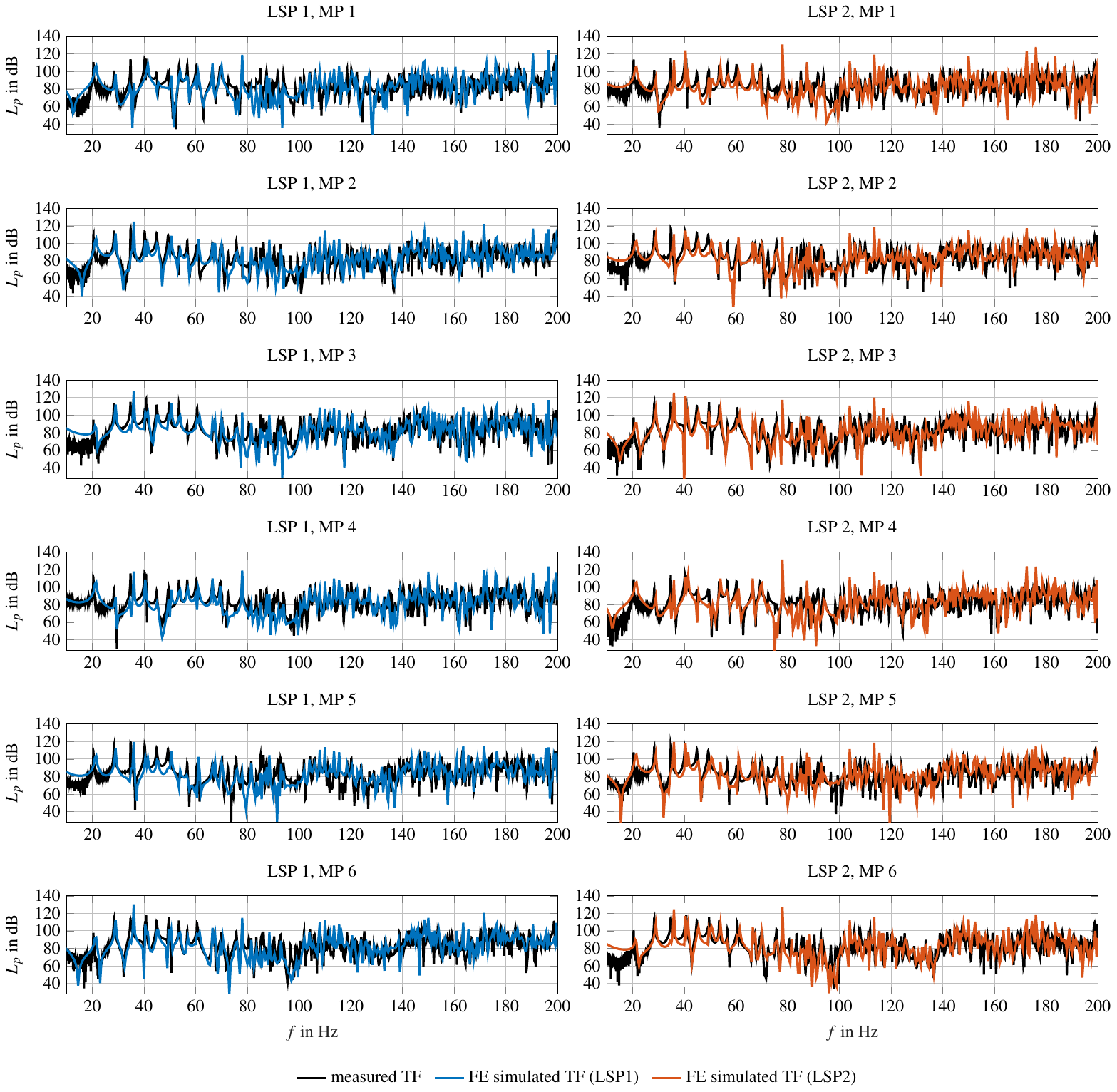}
    \caption{Comparison of measured and simulated TFs for the empty RC (without edge absorber). (LSP\dots loudspeaker position, MP\dots microphone position)}
    \label{fig:comp-TF-empty}
\end{figure}

\begin{figure}[htbp]
    \centering
    \includegraphics[width=\textwidth,keepaspectratio]{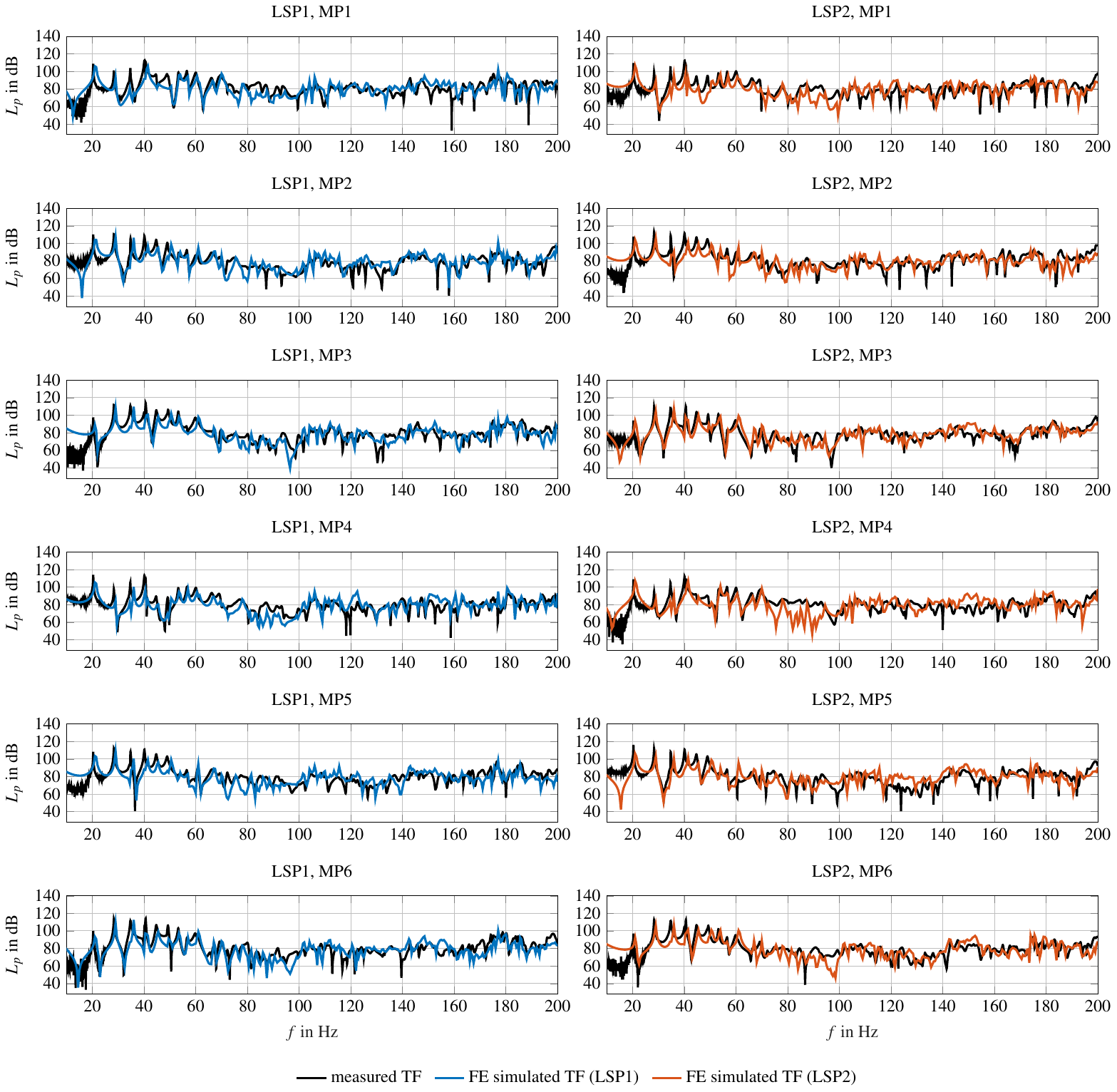}
    \caption{Comparison of measured and simulated TFs for edge absorber configuration EA1. (LSP\dots loudspeaker position, MP\dots microphone position)}
    \label{fig:comp-TF-config1}
\end{figure}

\begin{figure}[htbp]
	\centering
    \includegraphics[width=\textwidth,keepaspectratio]{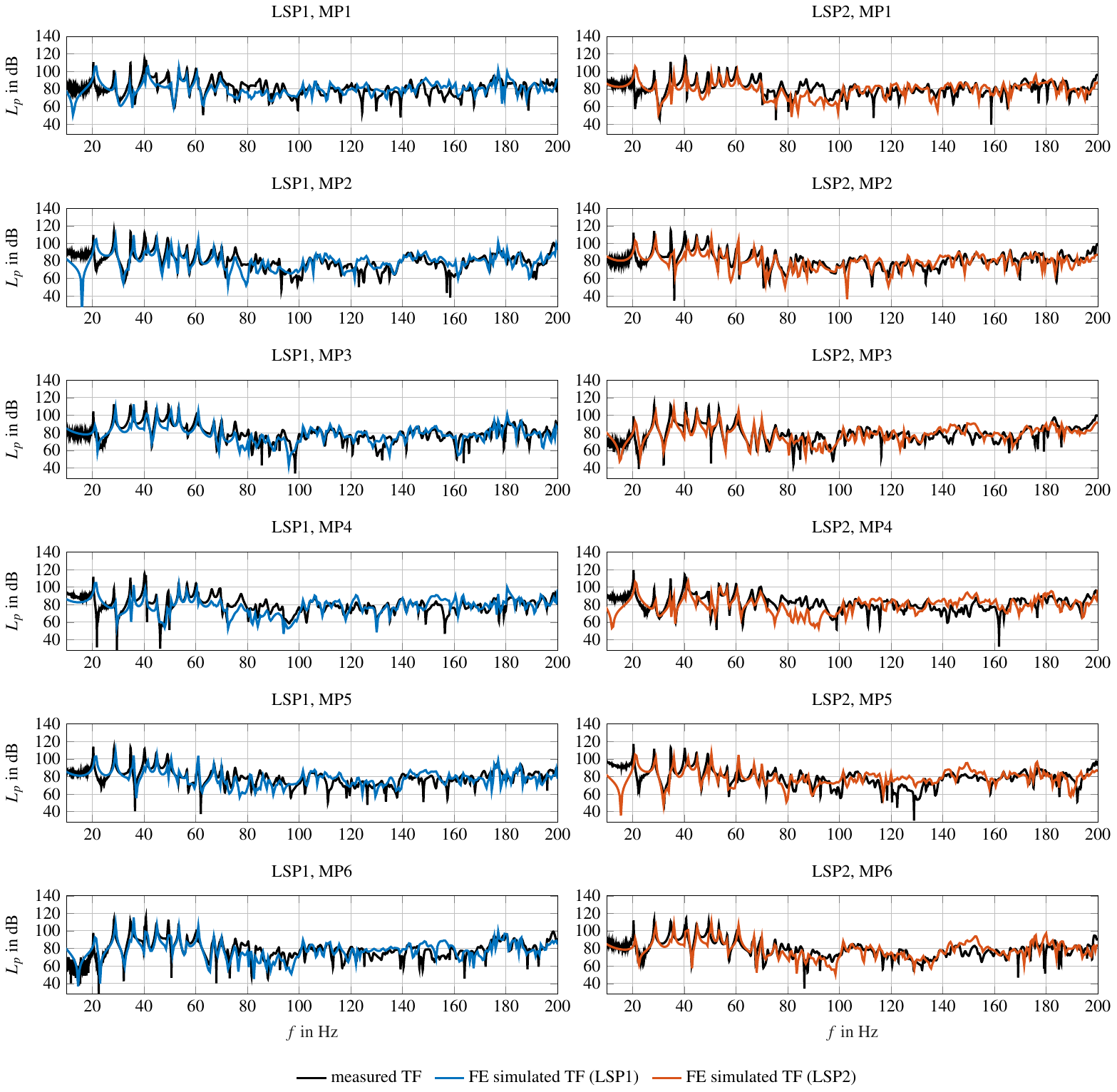}
	\caption{Comparison of measured and simulated TFs for edge absorber configuration EA2. (LSP\dots loudspeaker position, MP\dots microphone position)}
	\label{fig:comp-TF-config2}
\end{figure}

\begin{figure}[htbp]
	\centering
    \includegraphics[width=\textwidth,keepaspectratio]{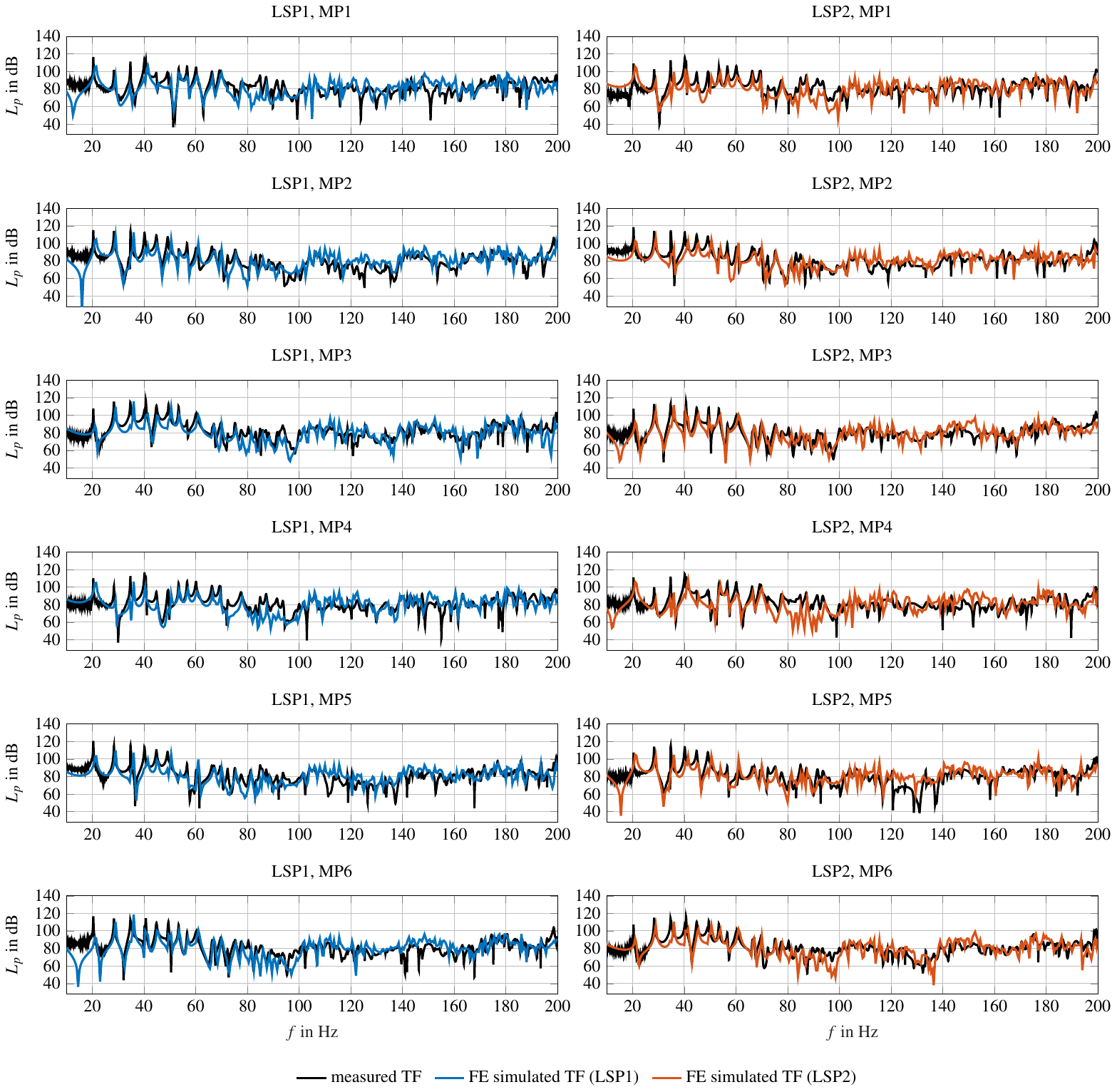}
	\caption{Comparison of measured and simulated TFs for edge absorber configuration EA3. (LSP\dots loudspeaker position, MP\dots microphone position)}
	\label{fig:comp-TF-config3}
\end{figure}

Figure \ref{fig:comp-TF-empty} depicts measured and simulated TFs for the empty room. In fig.~\ref{fig:comp-TF-config1}, measured and simulated TFs are compared for EA configuration EA1. Figure \ref{fig:comp-TF-config2} shows the TFs for the edge absorber configuration EA2. In fig.~\ref{fig:comp-TF-config3}, the TFs for edge absorber configuration EA3 are depicted.

\FloatBarrier
\clearpage
\printbibliography







\end{document}